\documentclass[paper]{JHEP3}
\usepackage{epsfig}
\usepackage{amssymb}
\def\beq{\begin{equation}}
\def\eeq{\end{equation}}

\def\beqn{\begin{eqnarray}}
\def\eeqn{\end{eqnarray}}

\def \as{\alpha_{\rm s}}

\def\HW{{\small HERWIG}}
\def\HWs{{\small HW6}}
\def\HWpp{{\small HW++}}
\def\ThePEG{{\small ThePEG}}
\def\NLO{{\small NLO}}
\def\MC{{\small MC}}
\def\EvG{{\small EvG}}

\def\yes{$\checkmark$}
\def\no{$\times$}
\newcommand\sss{\scriptscriptstyle\rm}

\newcommand\MCatNLO{{\rm MC}@{\rm NLO}}
\newcommand\code{\tt}
\newcommand\variable{\tt}
\newcommand\MSbar{{\overline {\rm MS}}}
\newcommand\sinthW{\sin\theta_{\sss W}}
\newcommand\sinsqthW{\sin^2\theta_{\sss W}}
\newcommand\sinfthW{\sin^4\theta_{\sss W}}
\newcommand\pt{p_{\sss T}}
\newcommand\amp{{\cal A}}
\preprint{
 Cavendish--HEP--10/12\hfill\\
 CERN-TH/2010-216\\
 IPPP/10/62\hfill\\
 DCPT/10/124\hfill
         }
\title{\boldmath The MC@NLO 4.0 Event Generator%
\footnote{Work supported in part by the UK Science and Technology 
Facilities Council, and by the Swiss National Science Foundation.}}
\author{Stefano Frixione%
  \thanks{On leave of absence from INFN, Sez. di Genova, Italy}\\
  PH Department, TH Unit, CERN, CH-1211 Geneva 23, Switzerland\\
  ITPP, EPFL, CH-1015 Lausanne, Switzerland\\
  E-mail: \email{Stefano.Frixione@cern.ch}}
\author{Fabian Stoeckli\\
  PH Department, CMG Group, CERN, CH-1211 Geneva 23, Switzerland\\
  E-mail: \email{Fabian.Stoeckli@cern.ch}}
\author{Paolo Torrielli\\
  ITPP, EPFL, CH-1015 Lausanne, Switzerland\\
  E-mail: \email{Paolo.Torrielli@epfl.ch}}
\author{Bryan R.\ Webber\\
  Cavendish Laboratory, 
  J.J. Thomson Avenue, Cambridge CB3 0HE, U.K.\\
  E-mail: \email{webber@hep.phy.cam.ac.uk}}
\author{Chris D. White\\
  Department of Physics and Astronomy, University of Glasgow,\\
  Glasgow G12 8QQ, Scotland, U.K.;\\
  Institute for Particle Physics Phenomenology, Department of Physics, 
  Durham University, Durham DH1 3LE, U.K.\\
  E-mail: \email{c.white@physics.gla.ac.uk}}
\abstract{
This is the user's manual of {$\MCatNLO$} 4.0. This package is a 
practical implementation, based upon the Fortran HERWIG and Herwig++ event 
generators, of the $\MCatNLO$ formalism, which allows one to incorporate 
NLO QCD matrix elements consistently into a parton shower framework.
Processes available in this version include the hadroproduction of
single vector and Higgs bosons, vector boson pairs, heavy
quark pairs, single top, single top in association with a $W$, 
single top in association with a charged Higgs in type I or II
2HDM models, lepton pairs, and Higgs bosons in association with a $W$ or $Z$. 
Spin correlations are included for all processes except $ZZ$ production. 
This document is self-contained, but we emphasise the main 
differences with respect to previous versions.
}
\begin{document}

\section{Generalities}
In this document, we briefly describe how to run the $\MCatNLO$ 
package, implemented according to the formalism introduced in 
ref.~\cite{Frixione:2002ik}. 
The production processes now available are listed in tables~\ref{tab:proc}
and~\ref{tab:procdec}. The process codes {\variable IPROC} and the
variables {\variable IV} and {\variable IL}$_\alpha$ will be explained 
below. $H_{1,2}$ represent hadrons (in practice, nucleons or antinucleons).
The information given in refs.~\cite{Frixione:2002ik,Frixione:2003ei}
allows the implementation in MC@NLO of any production process, provided
that the formalism of refs.~\cite{Frixione:1995ms,Frixione:1997np} is
used for the computation of cross sections to NLO accuracy. 
The production matrix elements have been taken from the following references:
vector boson pairs~\cite{Mele:1990bq,Frixione:1992pj,Frixione:1993yp},
$WZ$ production and decay with anomalous couplings~\cite{Dixon:1999di},
heavy quark pairs~\cite{Mangano:1991jk},
Standard Model Higgs~\cite{Dawson:1990zj,Djouadi:1991tk},
single vector boson~\cite{Altarelli:1979ub}, 
lepton pairs~\cite{Aurenche:1980tp},
associated Higgs~\cite{Oleari:2005inprep},
single-top $s$- and $t$-channel~\cite{Harris:2002md},
single-top in association with a charged Higgs~\cite{Weydert:2009vr};
those for single-top production in association with a $W$ have been 
re-derived and thoroughly compared to those of ref.~\cite{Giele:1995kr}.

This documentation refers to $\MCatNLO$ version 4.0. The major difference
with respect to previous $\MCatNLO$ versions is that in the present one
all processes can be simulated either with Fortran 
\HW~\cite{Marchesini:1992ch,Corcella:2001bw,Corcella:2002jc},
or with Herwig++~\cite{Bahr:2008pv,Bahr:2008tf} (with the 
exception, in the latter case, of single-top in association 
with a charged Higgs, and of $b\bar{b}$ production. While the
latter process is fully implemented at the level of short-distance 
cross sections, it displays an anomalously large event-failure rate 
in the shower phase, which is currently under investigation). 
In what follows, these two
event generators will be referred to as \HWs\ and \HWpp\ respectively,
and collectively as \EvG\footnote{The matching to PYTHIA has been
worked out  analytically in ref.~\cite{Torrielli:2010aw} for the case 
of initial-state radiation. The production processes implemented in computer
codes are still limited in number, and are not part of the present package.}.
Single-top production in association with a charged Higgs
has been added since version 3.4, including spin correlations which
were not yet implemented in ref.~\cite{Weydert:2009vr}. The present
version includes the upgrades of sub-versions 3.41 and 3.42. For precise 
details of version changes, see app.~\ref{app:newver}-\ref{app:newverh}.

\subsection{Citation policy\label{sec:cites}}
When using $\MCatNLO$ with \HWs, please cite ref.~\cite{Frixione:2002ik};
if \HWpp\ is adopted, please cite refs.~\cite{Frixione:2002ik,HWpppaper}.
In addition to ref.~\cite{Frixione:2002ik} or to 
refs.~\cite{Frixione:2002ik,HWpppaper}, if $t\bar{t}$ or $b\bar{b}$ events 
are generated, please also cite ref.~\cite{Frixione:2003ei}; if $s$- or 
$t$-channel single-top events are generated, please also cite 
ref.~\cite{Frixione:2005vw}; if $Wt$ single-top events are generated,
please also cite ref.~\cite{Frixione:2008yi}; if $H^\pm t$ single-top 
events are generated, please also cite ref.~\cite{Weydert:2009vr};
if $W^\pm Z$ events are generated, please also cite ref.~\cite{SFAOH}.
The current user manual, or any other user manuals relevant to past 
versions, should not be cited unless the relevant papers mentioned above 
are cited too.

\subsection{Physics processes\label{sec:oper}}
In the case of standard MC, a hard kinematic configuration is
generated on an event-by-event basis, and it is subsequently showered 
and hadronized. In the case of $\MCatNLO$, all of the hard kinematic
configurations are generated in advance, and stored in a file 
(which we call the {\em event file} -- see sect.~\ref{sec:evfile}); 
the event file is then read by the \EvG, which showers and 
hadronizes each hard configuration. Since version 2.0, the events are 
handled by the ``Les Houches'' generic user process 
interface~\cite{Boos:2001cv} (see ref.~\cite{Frixione:2003ei} for 
more details). Therefore, in $\MCatNLO$ the reading of a 
hard configuration from the event file is equivalent to the generation 
of such a configuration in a standard MC.

The signal to the \EvG\ that configurations should be read from an event file 
using the Les Houches interface is a negative value of the process code 
{\variable IPROC}; this accounts for the negative values in 
tables~\ref{tab:proc} and~\ref{tab:procdec}. 
In the case of heavy quark pair, Higgs, Higgs in association with a $W$ 
or $Z$, and lepton pair (through $Z/\gamma^*$ exchange) production, the codes 
are simply the negative of those for the corresponding standard \HWs\ MC
processes. Where possible, this convention will be adopted for additional
$\MCatNLO$ processes, regardless of the \EvG\ actually used to shower
events.  Consistently with what happens in standard \HWs, by
subtracting 10000 from {\variable IPROC} one generates the same processes as
in tables~\ref{tab:proc} and~\ref{tab:procdec}, but eliminates the underlying 
event\footnote{The same effect can be achieved by setting the \HWs\
parameter {\variable PRSOF} $=0$.}. Note that this operation is performed
within {\code mcatnlo\_hwdriver.f} (for \HWs), or by the scripts
{\code MCatNLO\_pp.Script} (for \HWpp). Therefore, if these files are
not employed when running the \EvG's, the user will be responsible for
switching the underlying event off manually if so desired.

For processes with a SM Higgs (denoted by $H^0$) in the final state,
the user may specify the identities of its decay products when using
\HWs, by adding {\variable -ID} to the process code. The conventions 
for {\variable ID} are the same as in \HWs, 
namely {\variable ID} $=1\ldots 6$ for 
$u\bar{u}\ldots t\bar{t}$; $7$, $8$, $9$ for $e^+e^-$, $\mu^+\mu^-$, 
$\tau^+\tau^-$; $10, 11$ for $W^+W^-, ZZ$; and 12 for $\gamma\gamma$. 
Furthermore, {\variable ID} $=0$ gives quarks of all flavours, and 
{\variable ID} $=99$ gives all decays. It should be stressed that 
the event file does not contain the Higgs decay products, and therefore
is independent of the value of {\variable ID}; the decay is
dealt with by \HWs.\footnote{Spin correlations between the decay 
 products of vector boson pairs emerging from Higgs decays 
 were neglected in \HWs\ versions older than the
 current one, 6.520. Please check the wiki at
  http://projects.hepforge.org/fherwig/trac/report
 for release reports on this and other improvements.}
In the case of \HWpp, the user will instead select the decay products
using the \HWpp\ input file {\code HWPPInput.inputs}, which we shall
describe later. This means that {\variable ID} has no effect in SM
Higgs production when using \HWpp.

Process codes {\variable IPROC}=$-1360-${\variable IL} and 
$-1370-${\variable IL} do not have an analogue in \HWs; they are the 
same as $-1350-${\variable IL}, except for the fact that only a $Z$ or a
$\gamma^*$ respectively is exchanged. The value of {\variable IL} determines
the lepton identities, and the same convention as in \HWs\ is adopted:
{\variable IL}=$1,\ldots,6$ for $l_{\rm IL}=e,\nu_e,\mu,\nu_\mu,\tau,\nu_\tau$
respectively (see also table~\ref{tab:ILval}).
At variance with \HWs, {\variable IL} cannot be set equal to
zero. Process codes {\variable IPROC}=$-1460-${\variable IL} and
$-1470-${\variable IL} are the analogue of \HWs\ $1450+${\variable IL};
in \HWs\ either $W^+$ or $W^-$ can be produced, whereas MC@NLO treats
the two vector bosons separately. For these processes, as in \HWs,
{\variable IL}=$1,2,3$ for $l_{\rm IL}=e,\mu,\tau$, but again
the choice ${\variable IL}=0$ is not allowed.

\begin{table}[h!]
\centering
\begin{tabular}{|c|c|c|c|c|l|}\hline
{\variable IPROC} & {\variable IV} & {\variable IL}$_1$ & {\variable IL}$_2$ & 
 Spin & Process \\\hline
 --1350--{\variable IL} & & & &\yes &
 $H_1 H_2\to (Z/\gamma^*\to) l_{\rm IL}\bar{l}_{\rm IL}+X$\\\hline
 --1360--{\variable IL} & & & &\yes &
 $H_1 H_2\to (Z\to) l_{\rm IL}\bar{l}_{\rm IL}+X$\\\hline
 --1370--{\variable IL} & & & &\yes &
 $H_1 H_2\to (\gamma^*\to) l_{\rm IL}\bar{l}_{\rm IL}+X$\\\hline
 --1460--{\variable IL} & & & &\yes &
 $H_1 H_2\to (W^+\to) l_{\rm IL}^+\nu_{\rm IL}+X$\\\hline
 --1470--{\variable IL} & & & &\yes &
 $H_1 H_2\to (W^-\to) l_{\rm IL}^-\bar{\nu}_{\rm IL}+X$\\\hline
 --1396 & & & &\no &
 $H_1 H_2\to \gamma^*(\to \sum_i f_i\bar{f}_i)+X$\\\hline
 --1397 & & & &\no &
 $H_1 H_2\to Z^0+X$\\\hline
 --1497 & & & &\no &
 $H_1 H_2\to W^+ +X$\\\hline
 --1498 & & & &\no &
 $H_1 H_2\to W^- +X$\\\hline
 --1600--{\variable ID} & & & & &
 $H_1 H_2\to H^0+X$\\\hline
 --1705 & & & & &
 $H_1 H_2\to b\bar{b}+X$\\\hline
 --1706 & & 7 & 7 & \no &
 $H_1 H_2\to t\bar{t}+X$\\\hline
 --2000--{\variable IC} & & 7 & & \no &
 $H_1 H_2\to t/\bar{t}+X$\\\hline
 --2001--{\variable IC} & & 7 & & \no &
 $H_1 H_2\to \bar{t}+X$\\\hline
 --2004--{\variable IC} & & 7 & & \no &
 $H_1 H_2\to t+X$\\\hline
 --2030 & & 7 & 7 & \no &
 $H_1 H_2\to tW^-/\bar{t}W^+ +X$\\\hline
 --2031 & & 7 & 7 & \no &
 $H_1 H_2\to \bar{t}W^+ +X$\\\hline
 --2034 & & 7 & 7 & \no &
 $H_1 H_2\to tW^- +X$\\\hline
 --2040 & & 7 & 7 & \no &
 $H_1 H_2\to tH^-/\bar{t}H^+ +X$\\\hline
 --2041 & & 7 & 7 & \no &
 $H_1 H_2\to \bar{t}H^+ +X$\\\hline
 --2044 & & 7 & 7 & \no &
 $H_1 H_2\to tH^- +X$\\\hline
 --2600--{\variable ID} & 1 & 7 & &\no &
 $H_1 H_2\to H^0 W^+ +X$\\\hline
 --2600--{\variable ID} & 1 & $i$ & &\yes &
 $H_1 H_2\to H^0 (W^+\to)l_i^+\nu_i +X$\\\hline
 --2600--{\variable ID} & -1 & 7 & &\no &
 $H_1 H_2\to H^0 W^- +X$\\\hline
 --2600--{\variable ID} & -1 & $i$ & &\yes &
 $H_1 H_2\to H^0 (W^-\to)l_i^-\bar{\nu}_i +X$\\\hline
 --2700--{\variable ID} & 0 & 7 & &\no &
 $H_1 H_2\to H^0 Z +X$\\\hline
 --2700--{\variable ID} & 0 & $i$ & &\yes &
 $H_1 H_2\to H^0 (Z\to)l_i\bar{l}_i +X$\\\hline
 --2850 & & 7 & 7 & \no &
 $H_1 H_2\to W^+W^-+X$\\\hline
 --2860 & & 7 & 7 & \no &
 $H_1 H_2\to Z^0Z^0+X$\\\hline
 --2870 & & 7 & 7 & \no &
 $H_1 H_2\to W^+Z^0+X$\\\hline
 --2880 & & 7 & 7 & \no &
 $H_1 H_2\to W^-Z^0+X$\\\hline
\end{tabular}
\caption{\label{tab:proc} 
Some of the processes implemented in $\MCatNLO~4.0$ (see also
table~\ref{tab:procdec}). $H_{1,2}$ represent nucleons or
antinucleons. {\variable IPROC}--10000 generates the same processes as 
{\variable IPROC}, but eliminates the underlying event. A void entry 
indicates that the corresponding variable is unused. The `Spin' column 
indicates whether spin correlations in vector boson or top decays are 
included (\yes), neglected (\no) or absent (void entry); when included,
spin correlations are obtained by direct integration of the relevant
NLO matrix elements. Spin correlations in Higgs decays to vector boson
pairs (e.g.\ $H^0\to W^+W^-\to l^+\nu l^-\bar{\nu}$) are included in
\HWs\ versions 6.520 and higher. Processes $-1705$ and $-2040-${\variable IC}
are not available for \HWpp\ at present.}
\end{table}

\begin{table}[h!]
\centering
\begin{tabular}{|c|c|c|c|c|l|}\hline
{\variable IPROC} & {\variable IV} & {\variable IL}$_1$ & {\variable IL}$_2$ & 
 Spin & Process \\\hline
 --1706 & & $i$ & $j$ & \yes &
 $H_1 H_2\to (t\to)b_k f_if^\prime_i 
             (\bar{t}\to)\bar{b}_l f_jf^\prime_j+X$\\\hline
 --2000--{\variable IC} & & $i$ & & \yes &
 $H_1 H_2\to (t\to)b_k f_if^\prime_i/
             (\bar{t}\to)\bar{b}_k f_if^\prime_i+X$\\\hline
 --2001--{\variable IC} & & $i$ & & \yes &
 $H_1 H_2\to (\bar{t}\to)\bar{b}_k f_if^\prime_i+X$\\\hline
 --2004--{\variable IC} & & $i$ & & \yes &
 $H_1 H_2\to (t\to)b_k f_if^\prime_i+X$\\\hline

 --2030 & & $i$ & $j$ & \yes &
 $H_1 H_2\to (t\to)b_k f_if^\prime_i (W^-\to)f_jf^\prime_j/$\\
        & & & & & 
 $\phantom{H_1 H_2\to\,}
 (\bar{t}\to)\bar{b}_k f_if^\prime_i(W^+\to)f_jf^\prime_j+X$\\\hline
 --2031 & & $i$ & $j$ & \yes &
 $H_1 H_2\to 
 (\bar{t}\to)\bar{b}_k f_if^\prime_i(W^+\to)f_jf^\prime_j+X$\\\hline
 --2034 & & $i$ & $j$ & \yes &
 $H_1 H_2\to (t\to)b_k f_if^\prime_i (W^-\to)f_jf^\prime_j+X$\\\hline
 --2040 & & $i$ &  & \yes &
 $H_1 H_2\to (t\to)b_k f_if^\prime_i H^-/$\\
        & & & & & 
 $\phantom{H_1 H_2\to\,}
 (\bar{t}\to)\bar{b}_k f_if^\prime_i H^+ +X$\\\hline
 --2041 & & $i$ &  & \yes &
 $H_1 H_2\to 
 (\bar{t}\to)\bar{b}_k f_if^\prime_i H^+ +X$\\\hline
 --2044 & & $i$ &  & \yes &
 $H_1 H_2\to (t\to)b_k f_if^\prime_i H^- +X$\\\hline
 --2850 & & $i$ & $j$ & \yes &
 $H_1 H_2\to (W^+\to)l_i^+\nu_i (W^-\to)l_j^-\bar{\nu}_j +X$\\\hline
 --2870 & & $i$ & $j$ & \yes &
 $H_1 H_2\to (W^+\to)l_i^+\nu_i (Z^0\to)l_j^\prime\bar{l}_j^\prime +X$\\\hline
 --2880 & & $i$ & $j$ & \yes &
 $H_1 H_2\to 
(W^+\to)l_i^-\bar{\nu}_i (Z^0\to)l_j^\prime\bar{l}_j^\prime +X$\\\hline
\end{tabular}
\caption{\label{tab:procdec} 
Some of the processes implemented in $\MCatNLO~4.0$ (see also
table~\ref{tab:proc}). $H_{1,2}$ represent nucleons or
antinucleons. For more details on $Wt$ and $H^\pm t$ production, see 
sect.~\ref{sec:Wt}. Spin correlations for the processes in this
table are implemented according to the method presented 
in ref.~\cite{Frixione:2007zp}. 
$b_\alpha$ ($\bar{b}_\alpha$) can either denote a $b$ (anti)quark
or a generic down-type (anti)quark. $f_\alpha$ and $f^\prime_\alpha$
can denote a (anti)lepton or an (anti)quark. See sects.~\ref{sec:xsecs} 
and~\ref{sec:decay} for fuller details. Process $-2040-${\variable IC}
is not available for \HWpp\ at present.}
\end{table}

The lepton pair processes {\variable IPROC}=$-1350-${\variable IL},
$\ldots$, $-1470-${\variable IL} include spin correlations when
generating the angular distributions of the
produced leptons. However, if spin correlations are not an
issue, the single vector boson production processes
{\variable IPROC}= $-$1396,$-$1397,$-$1497,$-$1498 can be used,
in which case the vector boson decay products are distributed
(by \HWs, which then generates the decays) according to phase space.
These processes should be considered only as a quicker alternative
to lepton-pair production, with a more limited physics content; for
this reason, they have not been interfaced to \HWpp.
There are a number of other differences between the lepton pair and single
vector boson processes. The latter do not feature the $\gamma$--$Z$
interference terms. Also, their cross sections are fully inclusive in the
final-state fermions resulting from $\gamma^*$, $Z$ or $W^\pm$.  The user can
still select a definite decay mode using the \HWs\ variable {\variable MODBOS} 
(see sect.~\ref{sec:decay}), but the relevant branching ratio will {\em not} be
included by MC@NLO. As stated previously, these processes are not available
for running with \HWpp.

In NLO computations for single-top production in the SM, it is customary 
to distinguish between three production mechanisms, conventionally
denoted as the $s$ channel, $t$ channel, and $Wt$ mode. Starting from 
version 3.4, all three mechanisms are implemented in $\MCatNLO$;
$s$- and $t$-channel single top production correspond to setting 
{\variable IC}=$10$ and {\variable IC}=$20$ respectively. 
For example, according to tables~\ref{tab:proc} and~\ref{tab:procdec}, 
$t$-channel single-$\bar{t}$ events will be generated by entering 
{\variable IPROC}=$-2021$. These two channels can also be simulated 
simultaneously (by setting {\variable IC}=$0$). We point out that the
$Wt$ cross section is ill-defined beyond the leading order in QCD,
which is also the case for $H^\pm t$ production when $m_H<m_t$.
See sect.~\ref{sec:Wt} for more details.

In the case of vector boson pair production, the process codes are the 
negative of those adopted in $\MCatNLO$ 1.0 (for which the Les Houches 
interface was not yet available), rather than those of standard \HWs.

Furthermore, in the case of $t\bar{t}$, single-$t$, $H^0W^\pm$, $H^0Z$, 
$W^+W^-$, and $W^\pm Z$ production, the value of {\variable IPROC} alone
may not be sufficient to fully determine the process type (including
decay products), and variables
{\variable IV}, {\variable IL}$_1$, and {\variable IL}$_2$ are also
needed (see tables~\ref{tab:proc} and~\ref{tab:procdec}).  
In the case of top decays (and of the decay of the hard $W$ 
in $Wt$ production), the variables {\variable IL}$_1$ 
and {\variable IL}$_2$ have a more extended range of values 
than that of the variable {\variable IL}, which is relevant to
lepton pair production and to which they are analogous
(notice, however, that in the latter case {\variable IL} is not 
an independent variable, and its value is included via {\variable
IPROC}). In addition, {\variable IL}$_\alpha$=7 implies that spin
correlations for the decay products of the corresponding particle are not
taken into account, as indicated in table~\ref{tab:proc}.
More details are given in sect.~\ref{sec:decay}.

Apart from the above differences, $\MCatNLO$ and \HWs\ or \HWpp\ 
{\em behave in exactly the same way}. Thus, the available user's 
analysis routines can be used in the case of $\MCatNLO$. One should recall, 
however, that $\MCatNLO$ always generates some events with negative weights 
(see ref.~\cite{Frixione:2002ik}); therefore, the correct distributions 
are obtained by summing weights with their signs (i.e., the absolute 
values of the weights must {\em NOT} be used when filling the histograms).

With such a structure, it is natural to create two separate executables,
which we improperly denote as \NLO\ and \MC. The former has the sole scope
of creating the event file; the latter is just the \EvG\ ``executable''. 
In the case of \HWs, this file coincides with the actual \HWs\ executable. 
In the case of \HWpp, it is a script that calls the relevant \HWpp\ 
executables.

\section{Structure of the package}
\subsection{Working environment}
We have written shell scripts and {\code Makefile}'s which will simplify 
the use of the package considerably. In order to use them, the computing system
must support {\code bash} shell, and {\code gmake}\footnote{For Macs 
running under OSX v10 or higher, {\code make} can be used instead of 
{\code gmake}.}. 
Should they be unavailable on the user's computing system, the compilation 
and running of $\MCatNLO$ requires more detailed instructions; in this case,
we refer the reader to app.~\ref{app:instr}. This appendix will serve also as
a reference for a more advanced use of the package.

\subsection{Source and running directories}
The package can be downloaded as a tarball from the web page:\\
$\phantom{aaaaaaaa}$%
{\code http://www.hep.phy.cam.ac.uk/theory/webber/MCatNLO}\\
The structure of the directories that contain the source codes has
become more involved starting from \MCatNLO\ version 4.0, in order to
deal with the possibility of interfacing to more than one \EvG.
The directory tree will be created automatically when unpacking
the tarball. We describe it here briefly.

The directory in which the tarball will be unpacked will be called
the {\em source directory}. The source directory contains the following 
files:
\begin{itemize}
\item[]
    {\code MCatNLO.inputs}\\
    {\code HWPPInput.inputs}\\
    {\code MCatNLO.Script}\\
    {\code MCatNLO\_pp.Script}\\
    {\code Makefile}\\
    {\code Makefile\_pp}\\
    {\code MCatNLO\_rb.inputs}
\end{itemize}
and the following subdirectories
\begin{itemize}
\item[]
    {\code HW6Analyzer}\\
    {\code HWppAnalyzer}\\
    {\code srcCommon}\\
    {\code srcHerwig6}\\
    {\code srcHerwigpp}\\
    {\code include}
\end{itemize}
The user will be primarily interested in {\code MCatNLO.inputs}
(or {\code MCatNLO\_rb.inputs}, see later), which contains 
all input parameters common to \HWs\ and to \HWpp. Further input 
parameters, specific to \HWpp, can be found in {\code HWPPInput.inputs}.
The user will have to write his/her own analysis routines, and place
them into the {\code HW6Analyzer} or the {\code HWppAnalyzer} directories
for \HWs\ or \HWpp\ runs respectively. Sample analysis files are provided
in order to give the user a ready-to-run package.
The other files of the package must not be modified, with the possible
exception of the \HWs\ driver ({\code mcatnlo\_hwdriver.f}),
and Les Houches interface ({\code mcatnlo\_hwlhin.f}), to be
found in {\code srcHerwig6} -- see sect.~\ref{sec:priors} for
further details.

When creating the executable, 
our shell scripts determine the type of operating system, and create a
subdirectory of the source directory, which we call the {\em running 
directory}, whose name depends on the operating system. For Linux,
the name of the running directory will be {\variable Linux} when
using \HWs, and {\variable LinuxPP} when using \HWpp. On other
operating systems, possible names of running directories are
{\variable AlphaXX}, {\variable SunXX}, or 
{\variable RunXX}, with {\variable XX} either the empty string
or {\variable PP}. Tests on operating systems other than Linux have 
been performed sporadically on Mac's, and in the case of \HWs\ only;
we recommend using Linux whenever possible.

The running directory contains all the object files and executable files, 
and in general all the files produced by the $\MCatNLO$ while running.  
It must also contain the relevant grid files (see sect.~\ref{sec:pdfs}), 
or links to them, if the library of parton densities provided with the 
$\MCatNLO$ package is used.

In the subdirectory {\code HW6Analyzer} of the source directory,
the user will find the files {\code mcatnlo\_hwan{\em xxx}.f} 
(which use a version of HBOOK written by M.~Mangano that outputs plots 
in TopDrawer format) and {\code mcatnlo\_hwan{\em xxx}\_rb.f} (which use 
front-end Fortran routines written by W.~Verkerke~\cite{WVroot} for filling 
histograms in Root format). These are sample \HWs\ analysis routines, one
for each of the processes implemented in this package. They are provided here 
to give the user a ready-to-run package, but they should be replaced 
with appropriate codes according to the user's needs. Examples of
how to use these analysis files in $\MCatNLO$ are given in the
(otherwise identical) {\code MCatNLO.inputs} and 
{\code MCatNLO\_rb.inputs} files (see sect.~\ref{sec:running}
for more details on input cards).
In the subdirectory {\code HWppAnalyzer} the user will find analogous
codes, to be used when running with \HWpp. We do not provide sample
analyses for all processes in this case, and only the TopDrawer-format
version is given, since Root is a native-C++ code that can be much
more easily used with \HWpp\ than with \HWs.

In addition to the files listed above, the user will need a version of 
the \EvG\ code(s) to be used -- 
\HWs~\cite{Marchesini:1992ch,Corcella:2001bw,Corcella:2002jc}
or \HWpp\cite{Bahr:2008pv,Bahr:2008tf}.
As stressed in ref.~\cite{Frixione:2002ik}, for $\MCatNLO$ we do not
modify the existing (LL) shower algorithm. However, since $\MCatNLO$
versions 2.0 and higher make use of the Les Houches interface,
first implemented in \HW\ 6.5, the version of \HWs\ must be 6.500 or higher.
When using \HWs, on most operating systems users will need to delete the 
dummy  subroutines {\small UPEVNT}, {\small UPINIT}, {\small PDFSET} and 
{\small STRUCTM} from the standard  \HWs\ package, to permit linkage of 
the corresponding routines from the $\MCatNLO$ package. As a general rule, 
the user is strongly advised to use the most recent versions of \HW,
which were the ones used in the testing phase of \MCatNLO.

\section{Prior to running\label{sec:priors}}

\subsection{Usage with \HWs}
When using \HWs, the user must be aware of the fact that the files:\\
$\phantom{aaa}${\code mcatnlo\_hwdriver.f}\\ 
$\phantom{aaa}${\code mcatnlo\_hwlhin.f}\\
which can be found in {\code srcHerwig6}, and the files:\\
$\phantom{aaa}${\code mcatnlo\_hwan{\em xxx}.f}\\
$\phantom{aaa}${\code mcatnlo\_hwan{\em xxx}\_rb.f}\\
which can be found in {\code HW6Analyzer}, 
contain the statement {\code INCLUDE HERWIG65.INC}, which indicates
that the code will link to version 6.500 or higher, for the
reasons explained above. In the current MC@NLO release, the file
{\code HERWIG65.INC} contains the statement\\
$\phantom{aaaaaa}${\code INCLUDE 'herwig6520.inc'}\\
We do not assume that the user will adopt version 6.520, which is
the latest release; for this reason, the user may need
to edit the file {\code HERWIG65.INC}, and change the statement
above into\\
$\phantom{aaaaaa}${\code INCLUDE 'herwig65nn.inc'}\\
with {\code 65nn} the version chosen by the user (this must be
consistent with the value of the input parameter {\variable HERWIGVER},
see sects.~\ref{sec:running} and~\ref{sec:scrvar}).

The file {\code mcatnlo\_hwdriver.f} contains a set of read statements,
which are necessary for \HWs\ to get the input parameters (see
sect.~\ref{sec:running} for the input procedure); these read
statements must not be modified or eliminated. Also, {\code
mcatnlo\_hwdriver.f} calls the \HWs\ routines which
perform showering, hadronization, decays (see sect.~\ref{sec:decay} 
for more details on this issue), and so forth; the user can
freely modify this part, as is customary in \HWs\ runs. Finally, the 
sample codes {\code mcatnlo\_hwan{\em xxx}.f} and
{\code mcatnlo\_hwan{\em xxx}\_rb.f} contain analysis-related routines:
these files must be replaced by files which contain the user's analysis 
routines. We point out that, since version 2.0, the {\code Makefile} need not
be edited any longer, since the corresponding operations are now 
performed by setting script variables (see sect.~\ref{sec:scrvar}).

\subsection{Usage with \HWpp}
When using \HWpp, no editing of the source codes is required,
except for those related to analysis, to be found in the directory
{\code HWppAnalyzer}. The analogues of the read statements of
\HWs\ {\code mcatnlo\_hwdriver.f} are, in the case of \HWpp, the
settings of parameters in {\tt HWPPInput.inputs}, which the user
can modify at will.
However, the user will have to provide installed versions of \HWpp\
and \ThePEG~\cite{Lonnblad:2006pt}. We advise the user to adopt
versions 2.4.2 for \HWpp\ and 1.6.1 for \ThePEG\ or later. 
We stress that the options {\tt ReconstructionOption=General} and 
{\tt InitialInitialBoostOption=LongTransBoost}~\\ must be used 
when running \HWpp. These are automatically set by our script,
and the user must not change them. The latter option is presently
available only in the trunk version of \HWpp\ 2.4.2.

\section{Running\label{sec:running}}
It is straightforward to run $\MCatNLO$. First, edit\footnote{See
below for comments on {\code MCatNLO\_rb.inputs}}\\
$\phantom{aaa}${\code MCatNLO.inputs}\\
and write there all the input parameters (for the complete list 
of the input parameters, see sect.~\ref{sec:scrvar}). Further parameters
specific to \HWpp\ that control the behaviour of this \EvG\ can be
set in {\tt HWPPInput.inputs}.
As the last line of the file {\code MCatNLO.inputs}, write\\
$\phantom{aaa}${\code runMCatNLO}\\
Finally, execute {\code MCatNLO.inputs} from the {\code bash} shell.
This procedure will create the \NLO\ and \MC\ executables, and run them
using the inputs given in {\code MCatNLO.inputs}, which guarantees
that the parameters used in the \NLO\ and \MC\ runs are consistent.
Should the user only need to create the executables without running
them, or to run the \NLO\ or the \MC\ only, he/she should replace the
call to {\code runMCatNLO} in the last line of {\code MCatNLO.inputs}
by calls to\\
$\phantom{aaa}${\code compileNLO}\\
$\phantom{aaa}${\code compileMC}\\
$\phantom{aaa}${\code runNLO}\\
$\phantom{aaa}${\code runMC}\\
which have obvious meanings. Take note that in the case of \HWpp, the
{\code compileMC} command actually does not compile the \HWpp\
executable. This has already been done in the installation of the
\HWpp\ package. The {\code compileMC} command rather compiles the chosen 
analyzer, creates all necessary soft links in the running directory, and 
creates the executable MC script. 

We point out that if using \HWs\ the command {\code runMC}
may be used with {\variable IPROC}=1350+{\variable IL}, 1450+{\variable IL}, 
1600+{\variable ID}, 1699, 1705, 1706, 2000--2008, 2600+{\variable ID}, 2699, 
2700+{\variable ID}, 2799, 2800, 2810, 2815, 2820, 2825 to generate 
$Z/\gamma^*$, $W^\pm$, Higgs, $b\bar{b}$, $t\bar{t}$, single top, $H^0W$,
$H^0Z$, and vector boson pair events with standard \HWs\ (see the \HWs\ 
manual for more details). Note that the events thus produced are 
weighted events (except for single-top production), since we have
set {\variable NOWGT=.FALSE.} in {\code mcatnlo\_hwdriver.f}; however, the
user can freely change this setting. For obvious reasons this does not 
work with \HWpp. In order to use standalone \HWpp\ please consult the 
dedicated manual.

We stress that the input parameters are not solely related to
physics (masses, CM energy, and so on); there are several of them
which control other things, such as the number of events generated.
These must also be set by the user, according to his/her needs:
see sect.~\ref{sec:scrvar}.

Two such variables are {\variable HERWIGVER} and {\variable HWUTI},
which were moved in version 2.0 from the {\code Makefile} to
{\code MCatNLO.inputs}, and which are relevant to runs with \HWs.
The former variable must be set equal to the 
object file name of the version of \HWs\ currently adopted 
(matching the one whose common blocks are included in the files
mentioned in sect.~\ref{sec:priors}). The variable {\variable HWUTI} 
must be set equal to the list of object files that the user needs in 
the analysis routines. In the case of \HWpp, the analogue of
{\variable HWUTI} is {\variable HWPPANALYZER}, which must be set
equal to the name of the C++ analysis code the user means to use.
The variable {\variable HERWIGVER} does not have an analogue when
using \HWpp; its role is played by the variables {\variable HWPPPATH}
and {\variable THEPEGPATH}, which must be set equal to the physical 
address of the base directories of the \HWpp, and of the \ThePEG\ 
installations, respectively.

The sample input file {\code MCatNLO.inputs} provided in this package 
is relevant to $t\bar{t}$ production and subsequent $t$ and $\bar{t}$  
leptonic decays. 
Similar sample inputs are given in the file {\code MCatNLO\_rb.inputs}, 
which is identical to the former, except that at the end of the MC run
an output file in Root format will be produced (as opposed to the
output file in TopDrawer format produced by {\code MCatNLO.inputs});
for this to happen, the user will have to edit {\code MCatNLO\_rb.inputs}
in order to insert the path to the Root libraries for the machine 
on which the run is performed (shell variables {\code EXTRAPATHS} and
{\code INCLUDEPATHS}). We stress that, apart from the differences in
the output formats, {\code MCatNLO.inputs} and {\code MCatNLO\_rb.inputs}
have exactly the same meaning. Thus, although for the sake of brevity
we shall often refer only to {\code MCatNLO.inputs} in this manual,
all the issues concerning the inputs apply to {\code MCatNLO\_rb.inputs}
as well. Furthermore, as stressed above, an explicit example of how
to obtain results in Root format is only necessary in the case of
\HWs\ runs, and therefore {\code MCatNLO\_rb.inputs} provided here
should be used only in conjunction with that \EvG.

If the shell scripts are not used to run the codes, the inputs are
given to the \NLO\ or \MC\ codes during an interactive talk-to phase;
the complete sets of inputs for our codes are reported in 
app.~\ref{app:input} for vector boson pair production.

\subsection{Parton densities\label{sec:pdfs}}
Since knowledge of the parton densities (PDFs) is necessary in
order to get the physical cross section, a PDF library must be
linked. For the NLO runs, the possibility exists to link the (now obsolete) 
CERNLIB PDF library (PDFLIB), or its replacement LHAPDF~\cite{Whalley:2005nh};
however, we also provide a self-contained PDF library with this package, 
which is faster than PDFLIB, and contains PDF sets released after the 
last and final PDFLIB version (8.04; most of these sets are now included 
in LHAPDF). 
The three PDF libraries mentioned above can also be used for
\HWs\ runs. As far as \HWpp\ is concerned, the only possibility is 
that of linking to LHAPDF, or that of using the default \HWpp\ PDF set. 
If this is desired, the variable {\variable HERPDF} must be set equal to 
{\code DEFAULT}~(see below for further comments).
A complete list of the PDFs available in our PDF library can 
be downloaded from the MC@NLO web page. The user may link one of the three
PDF libraries; all that is necessary is to set the variable {\variable
PDFLIBRARY} (in the file {\code MCatNLO.inputs}) equal to {\variable
THISLIB} if one wants to link to our PDF library, and equal to
{\variable PDFLIB} or to {\variable LHAPDF} if one wants to link 
to PDFLIB or to LHAPDF.  Our PDF library collects 
the original codes, written by the authors of the PDF fits;
as such, for most of the densities it needs to read the files which
contain the grids that initialize the PDFs. These files, which can
also be downloaded from the $\MCatNLO$ web page, must either be copied 
into the running directory, or defined in the running directory as logical
links to the physical files (by using {\code ln -sn}). We stress that if
the user runs $\MCatNLO$ with the shell scripts, the logical links will
be created automatically at run time. Starting from \MCatNLO\ version 3.4,
the reading of parameters associated with the selected PDF set from
LHAPDF has been made fully robust (see sect.~\ref{sec:lhapdf}). For
this reason, recent PDF sets are not being added to our PDF library,
and when adopting these (and/or when running \HWpp) LHAPDF must be used.

As stressed before, consistent inputs must be given to the \NLO\ and
\MC\ codes. However, in ref.~\cite{Frixione:2002ik} we found that the
dependence upon the PDFs used by \HWs\ is rather weak. So one may
want to run the \NLO\ and \HWs\ adopting a regular NLL-evolved set in the
former case, and the default \HWs\ set in the latter (the advantage is
that this option reduces the amount of running time of \HWs). In
order to do so, the user must set the variable {\variable HERPDF}
equal to {\variable DEFAULT} in the file {\code MCatNLO.inputs};
setting {\variable HERPDF=EXTPDF} will force \HWs\ to use the same
PDF set as the \NLO\ code.

On the other hand, we found that the use of the \HWpp\ default PDF
set can be problematic in versions where the default set is a 
so-called LO$^*$ PDF set. We observed unexpected features, especially 
in $\pt$ spectra. These features are manifest in both \MCatNLO\ and 
\HWpp\ standalone runs, and disappear when other sets are used.
We did not investigate further the origin of the problem, but we
deprecate the use of the \HWpp\ default PDF set with \MCatNLO\ and
recommend using the same set as in the \NLO\ run.

When using \HWs, regardless of the PDFs used in the \MC\ run users must 
delete the dummy PDFLIB routines {\small PDFSET} and {\small STRUCTM} 
from the \HWs\ source code, as explained earlier.

\subsubsection{LHAPDF\label{sec:lhapdf}}
As mentioned above, by setting {\variable PDFLIBRARY}$=${\variable LHAPDF}
in the input file the code is linked to the LHAPDF library. The user
may choose whether to link to the static or to the dynamic LHAPDF library
(the latter will produce a smaller executable but otherwise results are 
identical to those obtained with the former). This has obviously no effect 
for the MC step when using \HWpp, since the way LHAPDF is linked to the \HWpp\ 
executable is defined in the \HWpp\ installation. Starting from \MCatNLO\ 
version~4.0\footnote{In version 3.4 or earlier, linking to the static library
was the default, and linking to the dynamic one required the use
of the scripts in {\code MCatNLO\_dyn.Script}. These scripts are now
obsolete, and have been removed from the package, with their companion
{\code Makefile\_dyn}.}
this choice is made by assigning to the variable {\variable LHALINK} in
{\code MCatNLO.inputs} the values {\variable STATIC} or {\variable DYNAMIC}
respectively. In order for the {\code Makefile}'s to be able to find the 
LHAPDF library, the variable {\variable LHAPATH} in {\code MCatNLO.inputs} 
should be set equal to the name of the directory where the local 
version of LHAPDF is installed. This is typically the name of
the directory where one finds the files {\code libLHAPDF.a}
and {\code libLHAPDF.so}, except for the final {\code /lib} in
the directory name.

As is well known, a given PDF set has a preferred value of 
$\Lambda_{\sss QCD}$,
which should be used in the computation of short-distance cross sections.
Upon setting {\variable LAMBDAFIVE} in {\code MCatNLO.inputs} equal
to a negative value, this choice is made automatically. However, when
linking to PDFLIB or LHAPDF, the code has to rely on the value
$\Lambda_{\sss QCD}$ stored (by the PDF libraries) in a common block.
This is far from ideal, since $\Lambda_{\sss QCD}$ is not a physical
parameter, and in particular is dependent upon the form adopted for
$\as$, which may not be the same as that used in MC@NLO.
Starting from version 3.4, the above automatic choice has been
rendered more solid in the case of a linkage to LHAPDF; the code
now reads the value of $\as(M_Z)$ (i.e., of a physical quantity) 
from the PDF library, and converts it into a value for $\Lambda_{\sss QCD}$
using the form of $\as(Q^2)$ used internally in MC@NLO. MC@NLO will
print out on the standard output when running the NLO code
({\variable FPREFIXNLO.log} if using the scripts) the value of
$\Lambda_{\sss QCD}$ used in the computation. Such a value is now 
expected to be quite close to that listed under the column labeled
with $\Lambda_{\sss QCD}^{(5)}$(MeV) on our PDF library manual
(which can be found on the MC@NLO web page).

Version 4.0 of MC@NLO has been tested to link and run with several
versions of LHAPDF. In particular, the user is not supposed to edit 
the {\code Makefile}'s if linking with LHAPDF version 5.0 or higher.
If one is interested into linking with earlier versions of LHAPDF
(which is strongly deprecated), then one must replace the string 
{\code mcatnlo\_uti.o} in the variable {\variable LUTIFILES} in 
the {\code Makefile}, with the string {\code mcatnlo\_utilhav4.o}.
Again, the version of LHAPDF used in the MC step when using \HWpp\ is
defined when installing \HWpp. We advise however to try to always use
the same LHAPDF package, i.e. link \MCatNLO\ with the same library
as \HWpp.

\subsubsection{PDF uncertainties\label{sec:PDFerr}}
The use of error sets to estimate the uncertainties due to PDFs 
which affect cross sections implies one computation for each
of the members of the given error set -- when performing this
procedure with \MCatNLO, we recommend to set {\variable HERPDF=EXTPDF}.

The procedure is straightforward but computing intensive,
and an approximate solution is that of reweighting the results
obtained with the default PDF set by ratios of PDFs. In order
to do so, information is necessary on the values of the fractional 
momenta of the incoming partons $x_1$ and $x_2$, and on the scale squared 
$Q^2$ used in the computation of the PDFs. Starting from \MCatNLO\ 
version 4.0, this information is made available to the user
on an event-by-event basis, without the necessity of kinematical
reconstructions on the \EvG\ event record.
In the case of \HWs, the values of these quantities can be found
in the common block\\
$\phantom{aaaaaa}${\variable DOUBLE PRECISION UX1,UX2,UQ2}\\
$\phantom{aaaaaa}${\variable COMMON/CPDFRWGT/UX1,UX2,UQ2}\\
with {\variable UX1}$=x_1$, {\variable UX2}$=x_2$, and 
{\variable UQ2}$=Q^2$ (in GeV$^2$). In the case of \HWpp, the information
can be found right after the compulsory event information, in the form:\\
$\phantom{aaaaaa}${\variable \#pdf}~$\phantom{aa}x_1~x_2~Q^2$\\
We point out that reweighting with different PDFs is never exact in
the context of an NLO computation, and this is especially true when
such computation is interfaced to an event generator, as in \MCatNLO, 
since PDF effects in Sudakovs cannot possibly be taken into account
in this way. We therefore recommend performing PDF reweighting with
utmost care.

\subsection{Event file\label{sec:evfile}}
The \NLO\ code creates the event file. In order to do so, it goes through
two steps; first it integrates the cross sections (integration step),
and then, using the information gathered in the integration step, 
produces a set of hard events (event generation step). Integration and
event generation are performed with a modified version of the 
{\small SPRING-BASES} package~\cite{Kawabata:1995th}.

We stress that the events stored in the event file contain only the
partons involved in the hard subprocesses. Owing to the modified subtraction
introduced in the $\MCatNLO$ formalism (see ref.~\cite{Frixione:2002ik}) 
they do not correspond to pure NLO configurations, and should not be 
used to plot physical observables. Parton-level observables must be
reconstructed using the fully-showered events.

The event generation step necessarily follows the integration step;
however, for each integration step one can have an arbitrary number of
event generation steps, i.e., an arbitrary number of event files.
This is useful in the case in which the statistics accumulated 
with a given event file is not sufficient.

Suppose the user wants to create an event file; editing {\code
MCatNLO.inputs}, the user sets {\variable BASES=ON}, to enable the
integration step, sets the parameter {\variable NEVENTS} equal to
the number of events wanted on tape, and runs the code; the
information on the integration step (unreadable to the user, but
needed by the code in the event generation step) is written on files
whose name begin with {\variable FPREFIX}, a string the user sets
in {\code MCatNLO.inputs}; these files (which we denote as {\em data
files}) have extensions {\code .data}. The name of the event file is 
{\variable EVPREFIX.events}, where {\variable EVPREFIX} is again a 
string set by the user.

Now suppose the user wants to create another event file, to increase
the statistics. The user simply sets {\variable BASES=OFF}, since 
the integration step is not necessary any longer (however, the data
files must not be removed: the information
stored there is still used by the \NLO\ code); changes the string
{\variable EVPREFIX} (failure to do so overwrites the existing event
file), while keeping {\variable FPREFIX} at the same value as before;
and changes the value of {\variable RNDEVSEED} (the random number
seed used in the event generation step; failure to do so results in
an event file identical to the previous one); the number {\variable
NEVENTS} generated may or may not be equal to the one chosen in
generating the former event file(s).

We point out that data and event files may be very large. If the user
wants to store them in a scratch area, this can be done by setting the
script variable {\variable SCRTCH} equal to the physical address
of the scratch area (see sect.~\ref{sec:res}).

For historical reasons, the formats of the event files to be used
by \HWs\ and \HWpp\ are not identical; the former uses an internal
\MCatNLO\ format, whereas the latter uses the Les Houches format.
It is crucial to realize that event files meant to be showered by,
say, \HWpp\ must not be showered by \HWs\ (and the other way round).
This has nothing to do with the format of the event files (which
one may consider changing) and is instead due to the fact that the 
short-distance cross sections in \MCatNLO\ do depend on the \EvG\ 
used for the shower.

It has to be noted that in the case of \HWpp\ the command {\code runMC}
will typically not process the entire event file. The reason for this is 
the following. By using the Les Houches interface, a run will not stop
automatically when reaching the end of the event file, which will then
be oversampled. This happens since an \EvG\ counts as events processed
only those which are successfully showered, which are typically slightly
less than those read from the event file. In order to avoid oversampling,
we therefore give in input to \HWpp\ a number of events equal to 98\%
times {\variable NEVENTS} (the latter variable should coincide with
the number of events in the event file). The safety margin of 2\% is
amply sufficient to avoid oversampling. The user may change such a safety
margin by manipulating the MC executable script by hand, by changing the 
{\variable -N} parameter passed there to \HWpp.

\subsection{Inclusive NLO cross sections\label{sec:xsecs}}
MC@NLO integrates NLO matrix elements in order to produce the
event file, and thus computes (as a by-product) the inclusive
NLO cross section. This cross section (whose value is given
in $pb$) can be obtained from an MC@NLO run in three different 
ways when running \HWs\footnote{This is the case for items {\em b)}
and {\em c)} only if {\variable WGTTYPE}=1.}:
\begin{itemize}
\item[{\em a)}] It is printed out at the end of the \NLO\ run (search for
{\tt Total for fully inclusive} in the standard output).
\item[{\em b)}] It is printed by \HWs\ at the end of the \MC\ run
(search for {\tt CROSS SECTION (PB)} in the standard output).
\item[{\em c)}] It is equal to the integral of any differential 
distribution which covers the whole kinematically-accessible range
(e.g. $0\le\pt\le\infty$) and on which no cuts are applied.
\end{itemize}
These three numbers are the same ({\em up to statistics}, which here
means the number of generated events -- see the bottom of this section
for further comments) for the processes listed in table~\ref{tab:proc}.
For the processes listed in table~\ref{tab:procdec}, on the other
hand, the results of {\em b)} and {\em c)} are equal to
that of {\em a)}, times the branching ratio(s) for the selected
decay channel(s), times (in the case of top decays) other factors 
due to kinematic cuts specified in input (see below). This is so 
because for the processes of table~\ref{tab:procdec} spin correlations 
are obtained as described in ref.~\cite{Frixione:2007zp}. For these
processes, we shall denote in what follows the cross section obtained
in {\em a)} as the undecayed cross section, and those obtained in
{\em b)} or {\em c)} as the decayed cross sections. We note that,
both for the processes in table~\ref{tab:proc} and for those in
table~\ref{tab:procdec}, the results of {\em b)} and {\em c)} are 
equal to the sum of the weights of all events stored in the
event file (possibly up to the contributions of those few events
which \HWs\ is unable to shower and hadronize, and which are
therefore discarded with error messages in the \MC\ run).

The situation is basically identical in the case of \HWpp,
except for the fact that the result of {\em b)} is to be
found in the {\code *.out} file (in the running directory)
rather than in the standard output, and it is given in $nb$ 
rather than in $pb$.

For the processes of table~\ref{tab:procdec}, the branching ratios used 
in the computation are determined by the values of the branching ratios 
for individual decay channels. The following variables are relevant to 
top decays:
\beq
{\variable BRTOPTOLEP}=\frac{\Gamma\left(\sum_j t\to l\nu_l b_j\right)}
                            {\Gamma_t}\,,
\;\;\;\;\;\;\;\;
{\variable BRTOPTOHAD}=\frac{\Gamma\left(\sum_{ij} t\to u \bar{d}_i b_j\right)}
                            {\Gamma_t}\,,
\label{eq:BRtop}
\eeq
with $b_j$ and $\bar{d}_i$ any down-type quark and antiquark respectively,
$u$ an up-type quark, and $l$ a charged lepton; lepton and flavour universality
are assumed. In the case of $W$ decays, one has the analogous variables
\beq
{\variable BRWTOLEP}=\frac{\Gamma\left(W\to l\nu_l\right)}
                            {\Gamma_W}\,,
\;\;\;\;\;\;\;\;
{\variable BRWTOHAD}=\frac{\Gamma\left(\sum_i W\to u \bar{d}_i\right)}
                            {\Gamma_W}\,.
\label{eq:BRW}
\eeq
Finally, in the case of $Z$ decays one has
\beq
{\variable BRZTOEE}=\frac{\Gamma\left(Z\to l\bar{l}\right)}
                            {\Gamma_Z}\,.
\label{eq:BRZ}
\eeq
The variables in eqs.~(\ref{eq:BRtop})--(\ref{eq:BRZ}) can either be
given a numerical value in input, or computed at the LO in the SM by the
code -- see sect.~\ref{sec:decay} for details. The numerical values of
these variables are then combined to obtain the overall branching ratio
for the decay channels selected, which is done by setting the variables
{\variable IL}$_\alpha$ and {\variable TOPDECAY} as explained in 
sect.~\ref{sec:decay} (see in particular table~\ref{tab:ILval}).
For example, for a top decaying into a $W$ and any down-type 
quarks, with the $W$ decaying to an electron, muon, or any quarks, 
one sets {\variable IL}$_\alpha$=6, {\variable TOPDECAY=ALL},
and the resulting branching ratio will be 
$2\,\times\,${\variable BRTOPTOLEP}$+2\,\times\,${\variable BRTOPTOHAD}.

As mentioned above, in the case of top decays (as opposed to hard $W$ 
decays in $Wt$ or $W^+W^-$ or $W^\pm Z$ production) the decayed cross section
will include kinematic factors in addition to the branching ratios.
These factors are due to the fact that in general the range for the invariant
mass of the pair of particles emerging from the $W$ decay (i.e. the 
virtuality of the $W$) does not coincide with the maximum that is 
kinematically allowed. For each top that decays, the following 
kinematic factor will therefore be included in the decayed cross
section
\beq
\frac{\Gamma\left(t\to ff^\prime b\,|\,
q_{\sss W}({\rm inf}),q_{\sss W}({\rm sup})\right)}
{\Gamma\left(t\to ff^\prime b\,|\,0,m_t\right)}\,,
\eeq
with
\beq
\Gamma\left(t\to ff^\prime b\,|\,m,M\right)=
\int_{m^2}^{M^2} dq_{\sss W}^2\frac{d\Gamma\left(t\to ff^\prime b\right)}
{dq_{\sss W}^2}\,,
\eeq
and $q_{\sss W}({\rm inf})$, $q_{\sss W}({\rm sup})$ the lower and 
upper limits of the $W$ virtuality, which can be chosen in input.
In particular, if {\variable V1GAMMAX}$>0$, one will have
\beq
q_{\sss W}({\rm inf})={\variable WMASS}-
{\variable V1GAMMAX}\,\times\,{\variable WWIDTH}\,,
\;\;\;\;\;\;\;\;
q_{\sss W}({\rm sup})={\variable WMASS}+
{\variable V1GAMMAX}\,\times\,{\variable WWIDTH}\,.
\label{eq:rangeA}
\eeq
On the other hand, if {\variable V1GAMMAX} $<0$, one has
\beq
q_{\sss W}({\rm inf})={\variable V1MASSINF}\,,
\;\;\;\;\;\;\;\;
q_{\sss W}({\rm sup})={\variable V1MASSSUP}\,.
\label{eq:rangeB}
\eeq
The ranges in eqs.~(\ref{eq:rangeA}) or~(\ref{eq:rangeB}) apply
to the $W$ emerging from the decay of the top quark in $t\bar{t}$ production,
and of the top or antitop in single-top production (all channels). The
corresponding ranges for the $W$ emerging from the decay of the antitop 
quark in $t\bar{t}$ production are identical to those above, except for
the replacement of {\variable V1} with {\variable V2}.

The user is also allowed to generate events by fixing the virtuality of the 
$W$ emerging from top/antitop decays equal to the $W$ pole mass, by
setting {\variable xGAMMAX}$=0$, with {\variable x=V1,V2}. In such a
case, the decayed cross section will be equal to the undecayed cross
section, times the branching ratios, times a factor
\beq
\frac{d\Gamma\left(t\to ff^\prime b\right)}{dq_{\sss W}^2}
\Bigg|_{q_{\sss W}^2=M_{\sss W}^2}\,,
\eeq
for each decaying top quark. The decayed cross section will have 
therefore to be interpreted as differential in the $W$ virtuality squared
(doubly differential in the case of $t\bar{t}$ production), and
will be expressed in \mbox{$pb~{\rm GeV}^{-2}$} 
(or \mbox{$pb~{\rm GeV}^{-4}$} for $t\bar{t}$ production) units.

The branching ratios and kinematics factors for each decaying particle
are multiplied to give a single number (always less than or equal
to one), which is by definition the ratio of the decayed over the
undecayed cross section. This number is printed out at the end of the 
NLO run (search for {\tt Normalization factor due to decays} in the 
standard output).

We conclude this section by stressing that, 
while the result of {\em a)} is always computed with
a typical relative precision of $10^{-4}$, those of {\em b)} and
{\em c)} depend on the number of events generated. Although it has
been checked that, upon increasing the number of events generated, the
results of {\em b)} and {\em c)} do approach that of {\em a)} (possibly
times the branching ratios and kinematic factors), option {\em a)} 
is clearly preferred. As mentioned above, the decayed 
cross section of {\em b)} or {\em c)} can be obtained without
any loss of accuracy by multiplying the undecayed cross section of
{\em a)} by the normalization factor printed out by the code at
the end of the NLO run.

\subsection{$Wt$ and $H^\pm t$ production}\label{sec:Wt}
Owing to the interference with $t\bar{t}$ production, which occurs in 
the $gg$ and $q\bar{q}$ partonic channels starting at the NLO, the $Wt$ 
cross section is ill-defined beyond the leading order in QCD. One can 
still give an operative meaning to NLO $Wt$ production, but must
always be aware of the potential biases introduced in this way.
This issue and its potential physics implications are discussed
at length in ref.~\cite{Frixione:2008yi}, which the reader is
strongly advised to consult before generating $Wt$ events.

Starting from \MCatNLO\ version 3.4, we have implemented two different 
definitions of the $Wt$ cross section, which we denoted by 
{\em diagram removal} and {\em diagram subtraction} in 
ref.~\cite{Frixione:2008yi}.
The former computation is carried out by setting {\variable WTTYPE=REMOVAL}
in {\code MCatNLO.inputs}, while the latter corresponds to
{\variable WTTYPE=SUBTRACTION}. 
A practical application of these ideas to a phenomenological analysis 
is presented in ref.~\cite{White:2009yt}.

In $Wt$ production, the factorization (renormalization) scale is 
assigned the value of the variable {\variable PTVETO} (whose units
are GeV) if {\variable FFACT}$<0$ ({\variable FREN}$<0$). This option
should be used for testing purposes only; it is not recommended
in the generation of event samples for experimental studies.

In \MCatNLO\ version 4.0 we have implemented $H^\pm t$ production in
a generic 2HDM model (see ref.~\cite{Weydert:2009vr}). When $m_{H^\pm}<m_t$
this process interferes with $t\bar{t}$ production, and the same 
considerations as for $Wt$ production apply here. In order to avoid
the proliferation of input variables, {\variable WTTYPE} also controls
which definition of the $H^\pm t$ cross section is used.
When $m_{H^\pm}>m_t$ there is no interference with $t\bar{t}$ production,
and therefore $H^\pm t$ production is well defined and does not require
any special treatment at the level of matrix elements. In this mass
range, the user must set {\variable WTTYPE=REMOVAL} to run the code.
Finally, we point out that there are four input variables that
are specific to $H^\pm t$ production: {\variable TYPEIORII},
{\variable TANBETA}, {\variable ACPL}, and {\variable BCPL} --
see sect.~\ref{sec:scrvar} for more details.

\subsection{$W^\pm Z$ production and anomalous couplings}\label{sec:WZ}
Starting from \MCatNLO\ version 4.0, spin correlations have been added
to $W^\pm Z$ production. Furthermore, the user can now generate this
process either in the Standard Model, or with a Lagrangian with
anomalous (i.e., non-SM) couplings, according to ref.~\cite{Dixon:1999di}.
The most general amplitude for this process can be written as follows:
\beq
\amp=\amp_0+\Delta g_1^Z\amp_{\Delta g_1^Z}+
\Delta\kappa^Z\amp_{\Delta\kappa^Z}+\lambda^Z\amp_{\lambda^Z}\,,
\eeq
with $\amp_0$ the SM result. An event weight (i.e., the cross section)
will therefore be:
\beqn
w_{\sss TOT}&\propto&
w_0+2\Delta g_1^Z w_1 + 2\Delta\kappa^Z w_2 + 2\lambda^Z w_3
\nonumber\\*&&\phantom{w_0}
+2\Delta g_1^Z \Delta\kappa^Z w_4 +
2\Delta g_1^Z \lambda^Z w_5 +
2\Delta\kappa^Z \lambda^Z w_6 
\nonumber\\*&&\phantom{w_0}
+\left(\Delta g_1^Z\right)^2 w_7 + 
\left(\Delta\kappa^Z\right)^2 w_8 + 
\left(\lambda^Z\right)^2 w_9\,.
\label{anxsec}
\eeqn
The values of $\Delta g_1^Z$, $\Delta\kappa^Z$, and $\lambda^Z$
can be given in input using the script variables
{\variable DELG1Z}, {\variable DELKAPZ}, and {\variable LAMANZ}
respectively. By setting these three variables equal to zero 
one recovers the SM result. A fourth script variable, {\variable LAMFFAN},
corresponds to the quantity $\Lambda$ introduced in eq.~(8) of
ref.~\cite{Dixon:1999di}, and serves the purpose of avoiding violations
of unitarity; if it is set equal to zero in input, the program will
re-set it to the default value of 2~TeV.

Regardless of the values of $\Delta g_1^Z$, $\Delta\kappa^Z$ and $\lambda^Z$
given in input\footnote{We stress that this includes the SM case,
$\Delta g_1^Z=0$, $\Delta\kappa^Z=0$ and $\lambda^Z=0$.}, \MCatNLO\ can 
save the values of the weights $w_i$ of eq.~(\ref{anxsec}) in the event 
file; in order to do so, the user must set {\variable CPLWGT=YES} in 
the inputs. The quantities $w_i$ change event-by-event, and can be used
to re-weight the cross section and to obtain predictions simultaneously 
for any number of combinations of anomalous couplings. In the case of
\HWs, the information on $w_i$ is available in the MC run through
the common block\\
$\phantom{aaaaaa}${\variable DOUBLE PRECISION WGTACP(10)}\\
$\phantom{aaaaaa}${\variable COMMON/CWGTACP/WGTACP}\\
with {\variable WGTACP(I+1)}$=w_I/w_{\sss TOT}$. In the case of \HWpp, the 
information is available in the event file right before the 
{\variable </event>} tag, in the following form:\\
$\phantom{aaaaaa}${\variable \#an\_cpl\_wgt}~
$\phantom{aa}w_0/w_{\sss TOT}\,\ldots\, w_9/w_{\sss TOT}$

\subsection{Decays}\label{sec:decay}
$\MCatNLO$ is intended primarily for the study of NLO corrections
to production cross sections and distributions; NLO corrections to
the decays of produced particles are not included. As for spin 
correlations, the situation in version 4.0 is summarized 
in tables~\ref{tab:proc} and~\ref{tab:procdec}: they are included for all 
processes except $ZZ$ production\footnote{Non-factorizable 
spin correlations of virtual origin are not included in $W^+W^-$,
$t\bar{t}$, and single-$t$ production. See ref.~\cite{Frixione:2007zp}.}.
For the latter processes, quantities sensitive to the polarization of 
produced particles are not given correctly even to leading order.
For such quantities, it may be preferable to use the standard
\HW\ MCs, which do include leading-order spin correlations.

Following \HWs\ conventions, spin correlations in single-vector-boson
processes are automatically included using the process codes
({\variable IPROC}) relevant to lepton pair production (in other
words, if one is interested in including spin correlations in e.g. 
$W^+$ production and subsequent decays into $\mu^+\nu_\mu$, one needs to use
{\variable IPROC}$\,=\!-1461$ rather than {\variable IPROC}$\,=\!-1497$
and {\variable MODBOS}$(1)=3$). In order to avoid an 
unnecessary proliferation of {\variable IPROC} values, this strategy 
has not been adopted in other cases ($t\bar{t}$, single-$t$, $H^0W^\pm$, 
$H^0Z$, $W^+W^-$, $W^\pm Z$), in which spin correlations are included if
the variables {\variable IL}$_1$ and {\variable IL}$_2$ (the latter 
is used only in $t\bar{t}$, $Wt$, $W^+W^-$, and $W^\pm Z$ production) are 
assigned certain values. In the case of individual lepton decays,
these range from 1 to 3 if the decaying particle is a $W$ or a top,
or from 1 to 6 if the decaying particle is a $Z$. For these cases, the value of
{\variable IL}$_\alpha$ fully determines the identity of the leptons emerging
from the decay, and the same convention as in \HWs\ is adopted
(see the \HWs\ manual and sect.~\ref{sec:oper}).

In $t\bar{t}$ and single-top production, i.e. for all processes
listed in table~\ref{tab:procdec} which involve $t$ and/or $\bar{t}$, 
the top quark and/or antiquark,
and the hard $W$ in the case of $Wt$ production, can also decay hadronically.
In such cases, therefore, the variables {\variable IL}$_\alpha$ can
be assigned more values than for the other processes; the situation
is summarized in table~\ref{tab:ILval}. When generating the decays,
lepton and flavour universalities are assumed. The relative probabilities
of individual hadronic decays (e.g. $W^+\to u\bar{d}$ vs $W^+\to u\bar{s}$)
are determined using the CKM matrix elements entered by the user
(variables {\variable Vud} in {\code MCatNLO.inputs}). The relative
probabilities of leptonic vs hadronic decays are on the other hand
determined using the values of the corresponding branching ratios
entered by the user: variables {\variable BRTOPTOLEP} and
{\variable BRTOPTOHAD} for top/antitop decays, and {\variable BRWTOLEP}
and {\variable BRWTOHAD} for the decays of the hard $W$ emerging from
the hard process in $Wt$ production\footnote{{\variable BRWTOLEP} is
also used in $W^+W^-$ production. $W$ hadronic decays are not implemented
in this process, hence the branching ratio is only used as a rescaling
factor for event weights.} -- see eqs.~(\ref{eq:BRtop}) and~(\ref{eq:BRW})
for the definitions of these variables.
\begin{table}[htb]
\begin{center}
\begin{tabular}{|c|c|c|}\hline
{\variable IL}$_\alpha$ & $W$ decays & $Z$ decays \\
\hline\hline
0 & $e+\mu+\tau+q$ & \no \\\hline
1 & $e$            & $e^+e^-$ \\\hline
2 & $\mu$          & $\nu_e\bar{\nu}_e$ \\\hline
3 & $\tau$         & $\mu^+\mu^-$\\\hline
4 & $e+\mu$        & $\nu_\mu\bar{\nu}_\mu$ \\\hline
5 & $q$            & $\tau^+\tau^-$\\\hline
6 & $e+\mu+q$      & $\nu_\tau\bar{\nu}_\tau$ \\\hline
7 & no decay       & no decay \\\hline
\end{tabular}
\end{center}
\caption{\label{tab:ILval}
Decays of the $W$'s (produced in the hard process, or originating from 
top/antitop decays), and of the $Z$'s. The symbol $q$ denotes all hadronic
$W$ decays. Values different from 1, 2, or 3 for $W$ decays are only allowed 
in $t\bar{t}$ and single-top production (all channels).}
\end{table}

In the case of top/antitop decays, it is also possible to generate events 
in which the top decays into a $W$ and any down-type quark (hence the 
notations $b_\alpha$ and $\bar{b}_\alpha$ in table~\ref{tab:procdec}).
The identity of the latter is determined according to the CKM matrix 
values. For this to happen, one needs to set {\variable TOPDECAY=ALL}
in {\code MCatNLO.inputs}. If, on the other hand, one wants to
always generate $t\to Wb$ decays, one needs to set 
{\variable TOPDECAY=Wb}; in such a case, event weights (and thus
the decayed cross section, as defined in sect.~\ref{sec:xsecs})
will be multiplied by a factor $V_{tb}^2/(V_{td}^2+V_{ts}^2+V_{tb}^2)$.

In \MCatNLO\ version 4.0, spin correlations in leptonic processes involving 
intermediate $Z$ bosons are always included\footnote{In other words, 
hadronic $Z$ decays may be simulated by the \EvG, but spin correlations
are not included in such cases.} except in the case of $ZZ$ 
production. These processes are: dilepton production 
({\variable IPROC}$=-1350-${\variable IL}), $H^0Z$ production
({\variable IPROC}$=-2700-${\variable ID}), $W^+ Z$ production
({\variable IPROC}$=-2870$), and $W^- Z$ production
({\variable IPROC}$=-2880$).
In the case of dilepton and of $H^0Z$ production, the $Z$ may decay into 
charged-lepton or neutrino pairs. The identities of the decay products 
are determined by setting {\variable IL} (for dilepton production) or 
{\variable IL}$_1$ (for $H^0Z$ production) according to the values given 
in table~\ref{tab:ILval}. In the case of $W^\pm Z$ production, only
$Z$ decays into charged-lepton pairs are implemented, with the identities
of the leptons determined by setting the variable {\variable IL}$_2$
equal to 1, 3, or 5 (see table~\ref{tab:ILval}). 

For the processes in table~\ref{tab:procdec} it is also possible to 
force the code to use the LO values of the relevant leptonic and
hadronic branching ratios, by entering negative values for the top, 
$W$, and $Z$ widths (variables {\variable TWIDTH}, {\variable WWIDTH}, 
and {\variable ZWIDTH} in {\code MCatNLO.inputs}). In such a case,
the values of {\variable BRTOPTOLEP}, {\variable BRTOPTOHAD}, 
{\variable BRWTOLEP}, {\variable BRWTOHAD}, {\variable BRZTOEE} 
given in the input file will be ignored, and replaced by their LO
values (which are equal to $1/9$ for top and $W$ leptonic decays,
and equal to $1/3$ for top and $W$ hadronic decays). Likewise, the top, $W$,
and $Z$ widths will be computed using the LO SM formulae.

Spin correlations are implemented in the processes in table~\ref{tab:procdec} 
according to the method of ref.~\cite{Frixione:2007zp}, which is based
on a zero-width approximation for the decaying particles. Nevertheless,
the top quark and antiquark in $t\bar{t}$ production 
({\variable IPROC}$\,=\!-1706$), and the vector bosons in $W^+W^-$
({\variable IPROC}$\,=\!-2850$)  and $W^\pm Z$ 
({\variable IPROC}$\,=\!-2870/2880$) production can be given masses 
different from the pole masses. These off-shell effects are modeled
by re-weighting the cross section with skewed Breit-Wigner functions
(in order to take into account the fact that by changing the invariant
mass of the system produced one probes different values of Bjorken
$x$'s). This re-weighting is unitary, i.e. it does not change the
inclusive cross section. For $t\bar{t}$ production, the ranges
of top and antitop masses are controlled by the parameters
{\variable TiGAMMAX}, {\variable TiMASSINF}, and {\variable TiMASSSUP}
(with {\variable i}=1,2 for top and antitop respectively).
For $W^+W^-$ and $W^\pm Z$ production, one needs to use instead 
{\variable ViGAMMAX}, {\variable ViMASSINF}, and {\variable ViMASSSUP}, 
with {\variable i}=1,2 for $W^+$ and $W^-$, and for $W^\pm$ and $Z$ 
respectively. In both cases, the mass ranges will be defined by formulae 
formally identical to those of eqs.~(\ref{eq:rangeA}) and~(\ref{eq:rangeB}).
In version 4.0, off-shell effects are not implemented in the other
processes in table~\ref{tab:procdec}, i.e. all channels of
single-top production.

Finally, we point out that since spin correlations for the processes 
in table~\ref{tab:procdec} are implemented according to the method
of ref.~\cite{Frixione:2007zp}, tree-level matrix elements for
leptonic final states are needed. The codes for these have been generated
with MadGraph/MadEvent~\cite{Stelzer:1994ta,Maltoni:2002qb}, and
embedded into the MC@NLO package.

When {\variable IL}$_\alpha$=7, the corresponding particle is left undecayed 
by the \NLO\ code, and is passed as such to the \MC\ code; the information
on spin correlations is lost. However, the user can still force particular 
decay modes during the \MC\ run. When the chosen \EvG\ is \HWs, in the case 
of vector bosons, one proceeds in the same way as in standard \HWs, using 
the {\variable MODBOS}
variables -- see sect.~3.4 of ref.~\cite{Corcella:2001bw}. However,
top decays cannot be forced in this way because the decay is
treated as a three-body process: the $W^\pm$ boson entry in
{\code HEPEVT} is for information only.  Instead, the top
branching ratios can be altered using the {\variable HWMODK}
subroutine -- see sect.~7 of ref.~\cite{Corcella:2001bw}.
This is done separately for the $t$ and $\bar t$. For example,
{\code CALL HWMODK(6,1.D0,100,12,-11,5,0,0)} forces
the decay $t\to \nu_e e^+ b$, while leaving $\bar t$ decays
unaffected.  Note that the order of the decay products is
important for the decay matrix element
({\variable NME} = 100) to be applied correctly.
The relevant statements should be inserted in the \HW\ main program
(corresponding to {\code mcatnlo\_hwdriver.f} in this package)
after the statement {\code CALL HWUINC} and before
the loop over events.  A separate run with
{\code CALL HWMODK(-6,1.D0,100,-12,11,-5,0,0)} should
be performed if one wishes to symmetrize the forcing of
$t$ and $\bar t$ decays, since calls to {\variable HWMODK} from
within the event loop do not produce the desired result.
On the other hand, when \HWpp\ is used, the relevant decay channels
may be selected with suitable {\variable set} instructions, that
may be conveniently included in the {\code HWPPInput.inputs} file.
In the case a specific decay channel should be switched on/off, a
\mbox{{\code set /Herwig/Particles/}} command should be added, e.g.
\mbox{{\code set t->b,bbar,c;:OnOff Off}} to switch off the decay of 
top quarks into $b,\bar{b},c$. For more details please consult the 
\HWpp\ manual.

\subsection{Results\label{sec:res}}
As in the case of standard \HW\, the form of the results will be
determined by the user's analysis routines. However, in addition
to any files written by the user's analysis routines, the
$\MCatNLO$ writes the following files:\\
$\blacklozenge$ 
{\variable FPREFIXNLOinput}: the input file for the \NLO\ executable, 
created according to the set of input parameters defined in 
{\code MCatNLO.inputs} (where the user also sets the string
{\variable FPREFIX}). See table~\ref{tab:NLOi}.\\
$\blacklozenge$ 
{\variable FPREFIXNLO.log}: the log file relevant to the \NLO\ run.\\
$\blacklozenge$ 
{\variable FPREFIXxxx.data}: {\variable xxx} can assume several different 
values. These are the data files created by the \NLO\ code. They can be 
removed only if no further event generation step is foreseen with the
current choice of parameters.\\
$\blacklozenge$ 
{\variable FPREFIXMCinput}: analogous to {\variable FPREFIXNLOinput}, 
but for the \MC\ executable. See table~\ref{tab:MCi}.\\
$\blacklozenge$ 
{\variable FPREFIXMC.log}: analogous to {\variable FPREFIXNLO.log}, but 
for the \MC\ run.\\
$\blacklozenge$ 
{\variable FPREFIXMC.out}: produced only in \HWpp\ runs. Contains 
the result for the inclusive NLO rate (see sect.~\ref{sec:xsecs})\\
$\blacklozenge$ 
{\variable EVPREFIX.events}: the event file, where {\variable EVPREFIX} 
is the string set by the user in {\code MCatNLO.inputs}.\\
$\blacklozenge$ 
{\variable EVPREFIXxxx.events}: {\variable xxx} can assume several different 
values. These files are temporary event files, which are used by the
\NLO\ code, and eventually removed by the shell scripts. They MUST NOT be
removed by the user during the run (the program will crash or give
meaningless results).

In addition to the above one will get the following files when running
with \HWpp:\\
$\blacklozenge$ 
{\variable FPREFIXMC.run}: This file is produced by the \HWpp\
executable and is internally needed for the run. It is in simplified
words the compiled {\variable FPREFIXMCinput} file, resulting from the
{\code Herwig++ read FPREFIXMCinput} command, which is executed in the
executable script.\\
$\blacklozenge$ 
{\variable FPREFIXMC.tex}: Information, including references,
describing the settings used in the \HWpp\ run in \LaTeX\ format.

By default, all the files produced by the $\MCatNLO$ are written in the
running directory.  However, if the variable {\variable SCRTCH} (to be set in
{\code MCatNLO.inputs}) is {\em not} blank, the data and event files will be
written in the directory whose address is stored in {\variable SCRTCH}
(such a directory is not created by the scripts, and must already exist
at run time).

\section{Script variables\label{sec:scrvar}}
In the following, we list all the variables appearing in 
{\code MCatNLO.inputs}; these can be changed by the user to suit 
his/her needs. This must be done by editing {\code MCatNLO.inputs}.
For fuller details see the comments in {\code MCatNLO.inputs}.
\begin{itemize}
\item[{\variable ECM}] 
 The CM energy (in GeV) of the colliding particles.
\item[{\variable FREN}] 
 The ratio between the renormalization scale, and a reference mass scale.
\item[{\variable FFACT}] 
 As {\variable FREN}, for the factorization scale.
\item[{\variable HVQMASS}] 
 The mass (in GeV) of the top quark, except when 
 {\variable IPROC}$\,=\!-(1)1705$,
 when it is the mass of the bottom quark. In this case, {\variable HVQMASS} 
 must coincide with {\variable BMASS}.
\item[{\variable xMASS}] 
 The mass (in GeV) of the particle {\variable x}, with 
 {\variable x=HGG,W,Z,U,D,S,C,B,G}.
\item[{\variable xWIDTH}] 
 The physical (Breit-Wigner) width (in GeV) of the particle {\variable x}, 
 with {\variable x=HGG,W,Z,T} for $H^0$, $W^\pm$, $Z$, and $t$ respectively.
\item[{\variable BRTOPTOx}] 
 Branching ratio for top decay channels $\sum_j t\to l\nu_l b_j$ (when
 {\variable x=LEP}) and $\sum_{ij} t\to u \bar{d}_i b_j$ (when
 {\variable x=HAD}). Lepton and flavour universality is assumed.
\item[{\variable BRWTOx}] 
 Branching ratio for $W$ decay channels $W\to l\nu_l$ (when
 {\variable x=LEP}) and $\sum_i W\to u \bar{d}_i$ (when
 {\variable x=HAD}). Lepton and flavour universality is assumed.
\item[{\variable BRZTOEE}] 
 Branching ratio for $Z$ decay channels $Z\to l\bar{l}$.
 Lepton universality is assumed.
\item[{\variable IBORNHGG}] 
 Valid entries are 1 and 2.  If set to 1, the exact top mass dependence is
 retained {\em at the Born level} in Higgs production.  If set to 2, the
 $m_t\to\infty$ limit is used.
\item[{\variable xGAMMAX}] 
 If {\variable xGAMMAX} $>0$, controls the width of the mass range for 
 Higgs ({\variable x=H}), vector bosons ({\variable x=V1,V2}), and
 top ({\variable x=T1,T2}): the range is 
 ${\variable MASS}\pm({\variable GAMMAX} \times{\variable WIDTH})$.
 Off-shell effects for top are only implemented in $t\bar{t}$ production.
\item[{\variable xMASSINF}] 
 Lower limit of the Higgs ({\variable x=H}), vector boson
 ({\variable x=V1,V2}), and top ({\variable x=T1,T2})
 mass range; used only when {\variable xGAMMAX} $<0$.
\item[{\variable xMASSSUP}] 
 Upper limit of the Higgs ({\variable x=H}), vector boson 
 ({\variable x=V1,V2}), and top ({\variable x=T1,T2})
 mass range; used only when {\variable xGAMMAX} $<0$.
\item[{\variable Vud}]
 CKM matrix elements, with {\variable u}={\variable U,C,T} and
 {\variable d}={\variable D,S,B}. Set {\variable VUD=VUS=VUB}=0
 to use values of PDG2003. 
\item[{\variable AEMRUN}]
 Set it to {\variable YES} to use running $\alpha_{em}$ in lepton pair and
 single vector boson production, set it to {\variable NO} to use 
 $\alpha_{em}=1/137.0359895$.
\item[{\variable TYPEIORII}]
 Set this variable equal to 1 or 2 in order to use a type-I or a type-II
 2HDM model for the computation of $H^\pm t$ production.
\item[{\variable TANBETA}]
 Set this variables equal to $\tan\beta$; effective only for $H^\pm t$ 
 production in the context of a type-II 2HDM model.
\item[{\variable xCPL}]
 Here, {\variable x=A,B}. Set these variables equal to the $A$ (scalar) and
 $B$ (pseudoscalar) coefficients of the $tHb$ vertex, in a type-I 2HDM model.
 Effective only for $H^\pm t$ production.
\item[{\variable IPROC}]
 Process number that identifies the hard subprocess: see tables~\ref{tab:proc} 
 and~\ref{tab:procdec} for valid entries.
\item[{\variable IVCODE}]
 Identifies the nature of the vector boson in associated Higgs production.
 It corresponds to variable {\variable IV} of table~\ref{tab:proc}.
\item[{\variable ILxCODE}]
 Identify the nature of the particles emerging from vector boson or top 
 decays. They correspond to variables {\variable IL}$_1$ and 
 {\variable IL}$_2$ (for {\variable x} $=1,2$ respectively) of 
 tables~\ref{tab:proc}, \ref{tab:procdec} and~\ref{tab:ILval}.
\item[{\variable TOPDECAY}]
 Valid entries are {\variable ALL} and {\variable Wb}. Controls the type
 of top decay. See sect.~\ref{sec:decay}.
\item[{\variable WTTYPE}]
 Valid entries are {\variable REMOVAL} and {\variable SUBTRACTION}. Determines
 the definition of the $Wt$ and $H^\pm t$ cross sections at the NLO. See 
 sect.~\ref{sec:Wt}.
\item[{\variable PTVETO}]
 Used in conjunction with {\variable FFACT} and/or {\variable FREN} to 
 set mass scales in $Wt$ production. See sect.~\ref{sec:Wt}.
\item[{\variable PARTn}]
 The type of the incoming particle \#{\variable n}, with {\variable n}=1,2. 
 \HWs\ naming conventions are used ({\variable P, PBAR, N, NBAR}).
\item[{\variable PDFGROUP}]
 The name of the group fitting the parton densities used;
 the labeling conventions of PDFLIB are adopted. Unused when linked
 to LHAPDF.
\item[{\variable PDFSET}] 
 The number of the parton density set; according to PFDLIB conventions,
 the pair ({\variable PDFGROUP}, {\variable PDFSET}) identifies the 
 densities for a given particle type. When linked to LHAPDF, use 
 the numbering conventions of LHAGLUE~\cite{Whalley:2005nh}.
\item[{\variable LAMBDAFIVE}]
 The value of $\Lambda_{\sss QCD}$, for five flavours and in the 
 ${\overline {\rm MS}}$ scheme, used in the computation of NLO
 cross sections. A negative entry sets $\Lambda_{\sss QCD}$ equal
 to that associated with the PDF set being used.
\item[{\variable LAMBDAHERW}]
 The value of $\Lambda_{\sss QCD}$ used in MC runs; this parameter has the 
 same meaning as $\Lambda_{\sss QCD}$ in \HW.
\item[{\variable SCHEMEOFPDF}] 
 The subtraction scheme in which the parton densities are defined.
\item[{\variable FPREFIX}] Our integration routine creates files with
 name beginning by the string {\variable FPREFIX}. Most of these files are not 
 directly accessed by the user. See sects.~\ref{sec:evfile} 
 and~\ref{sec:res}.
\item[{\variable EVPREFIX}] 
 The name of the event file begins with this string. See 
 sects.~\ref{sec:evfile} and~\ref{sec:res}.
\item[{\variable EXEPREFIX}] 
 The names of the \NLO\ and \MC\ executables begin with this string; this is
 useful in the case of simultaneous runs.
\item[{\variable NEVENTS}] 
 The number of events stored in the event file, eventually
 processed by the \EvG. See sect.~\ref{sec:evfile} for comments
 relevant to \HWpp.
\item[{\variable MCMODE}] 
 Valid entries are {\variable HW6} and {\variable HWPP}, for using 
 \HWs\ or \HWpp\ in the \MC\ run respectively.
\item[{\variable WGTTYPE}]
 Valid entries are 0 and 1. When set to 0, the weights in the event file
 are $\pm 1$. When set to 1, they are $\pm w$, with $w$ a constant such
 that the sum of the weights gives the total inclusive NLO cross section
 (see sect.~\ref{sec:xsecs} for more details).
 Note that these weights are redefined by the \EvG\ at \MC\ run time 
 according to its own convention (see \HWs\ or \HWpp\ manual).
\item[{\variable RNDEVSEED}] 
 The seed for the random number generation in the
 event generation step; must be changed in order to obtain
 statistically-equivalent but different event files.
\item[{\variable BASES}] 
 Controls the integration step; valid entries are {\variable ON} and 
 {\variable OFF}. At least one run with {\variable BASES=ON} must be 
 performed (see sect.~\ref{sec:evfile}).
\item[{\variable PDFLIBRARY}] 
 Valid entries are {\variable PDFLIB}, {\variable LHAPDF}, and 
 {\variable THISLIB}. In the former two cases, PDFLIB or LHAPDF is used to 
 compute the parton densities, whereas in the latter case the densities are
 obtained from our self-contained PDF library.
\item[{\variable HERPDF}] 
 If set to {\variable DEFAULT}, the \EvG\ uses its internal PDF set 
 (controlled by {\variable NSTRU} in the case of \HWs), regardless of the 
 densities adopted at the NLO level. If set to {\variable EXTPDF}, the 
 \EvG\ uses the same PDFs as the \NLO\ code (see sect.~\ref{sec:pdfs}).
\item[{\variable HWPATH}]
 The physical address of the directory where the user's
 preferred version of \HWs\ is stored.
\item[{\variable HWPPPATH}]
 The physical address of the directory where the user's
 preferred version of the \HWpp\ tool is installed. Point to the base
 directory that holds the {\code lib, bin, include} and {\code share}
 directories. This is typically the directory one has specified in
 the {\code configure} step of the \HWpp\ installation with the {\code
 --prefix} parameter. 
\item[{\variable THEPEGPATH}]
 The analogue of {\variable HWPPPATH}, for the \ThePEG\ library.
Point to the base
 directory that holds the {\code lib, bin, include} and {\code share}
 directories. This is typically the directory you have specified in
 the {\code configure} step of the \ThePEG\ installation with the {\code
 --prefix} parameter. 
\item[{\variable SCRTCH}]
 The physical address of the directory where the user wants to store the
 data and event files. If left blank, these files are stored in the 
 running directory.
\item[{\variable HWUTI}]
 This variables must be set equal to a list of object files,
 needed by the \HWs\ analysis routines of the user (for example,
 {\variable HWUTI=''obj1.o obj2.o obj3.o''} is a valid assignment).
\item[{\variable HWPPANALYZER}]
 This variables must be set equal to the name of the C++ analysis
 file relevant to \HWpp\ runs (for example, 
 {\variable HWPPANALYZER=''TopAnalysis''} is a valid assignment, with
 {\code TopAnalysis.cc} and {\code TopAnalysis.h} being files in
 the {\code HWppAnalyzer} directory).
\item[{\variable HERWIGVER}]
 This variable must to be set equal to the name of the
 object file corresponding to the version of \HWs\ linked
 to the package (for example, {\variable HERWIGVER=''herwig6520.o''} is a
 valid assignment).
\item[{\variable PDFPATH}]
 The physical address of the directory where the PDF grids are stored.
 Effective only if {\variable PDFLIBRARY=THISLIB}.
\item[{\variable LHALINK}]
 Set this variable equal to {\variable STATIC} or {\variable DYNAMIC}
 for linking with the static or dynamic LHAPDF library with \HWs. 
 This variable has no effect on how the \HWpp\ executable is linked 
 with LHAPDF.
\item[{\variable LHAPATH}]
 Set this variable equal to the name of the directory where the local 
 version of LHAPDF is installed. See sect.~\ref{sec:lhapdf}. 
 This has no effect on which version of LHAPDF is used in the MC step
 when running \HWpp. 
\item[{\variable LHAOFL}]
 Set {\variable LHAOFL=FREEZE} to freeze PDFs from LHAPDF at the boundaries,
 or equal to {\variable EXTRAPOLATE} otherwise. See LHAPDF manual for
 details.
\item[{\variable EXTRALIBS}]
 Set this variable equal to the names of the libraries which need be linked.
 LHAPDF is a special case, and must not be included in this list.
\item[{\variable EXTRAPATHS}]
 Set this variable equal to the names of the directories where the 
 libraries which need be linked are installed.
\item[{\variable INCLUDEPATHS}]
 Set this variable equal to the names of the directories which 
 contain header files possibly needed by C++ files provided by the user
 (also in \HWs\ runs, via {\variable HWUTI}). 
\end{itemize}

\section*{Acknowledgments}
It is a pleasure to thank the co-authors of the MC@NLO papers,
E.~Laenen, P.~Motylinski, and P.~Nason, for having contributed 
so much to many different aspects of the $\MCatNLO$ project,
and for stimulating discussions. We thank W.~Verkerke for having 
provided us with a Fortran interface to C++ Root-calling routines.
We thank F.~Filthaut for having uncovered and fixed a bug
in version 3.4. B.R.~W. thanks the CERN theory group for frequent hospitality.
The work of P.~T. is supported by the Swiss National Science Foundation.
C.D.~W. is supported by the STFC Postdoctoral Fellowship ``Collider 
Physics at the LHC''.
Finally, we are indebted with all the members of experimental 
collaborations, unfortunately too numerous to be explicitly mentioned 
here, who used the code and gave us precious suggestions and feedback.

\section*{Appendices}
\appendix

\section{Version changes}

\subsection{From MC@NLO version 1.0 to version 2.0\label{app:newver}}
In this appendix we list the changes that occurred in the package
from version 1.0 to version 2.0.

$\bullet$~The Les Houches generic user process interface has been adopted.

$\bullet$~As a result, the convention for process codes has been changed:
MC@NLO process codes {\variable IPROC} are negative.

$\bullet$~The code {\code mcatnlo\_hwhvvj.f}, which was specific to
vector boson pair production in version 1.0, has been replaced by
{\code mcatnlo\_hwlhin.f}, which reads the event file according
to the Les Houches prescription, and works for all the production 
processes implemented.

$\bullet$~The {\code Makefile} need not be edited, since the variables
{\variable HERWIGVER} and {\variable HWUTI} have been moved to
{\code MCatNLO.inputs} (where they must be set by the user).

$\bullet$~A code {\code mcatnlo\_hbook.f} has been added to the list of
utility codes. It contains a simplified version (written by M.~Mangano)
of {\small HBOOK}, and it is only used by the sample analysis routines
{\code mcatnlo\_hwan{\em xxx}.f}. As such, the user will not need it
when linking to a self-contained analysis code.

We also remind the reader that the \HWs\ version must be 
6.5 or higher since the  Les Houches interface is used.

\subsection{From MC@NLO version 2.0 to version 2.1\label{app:newvera}}
In this appendix we list the changes that occurred in the package
from version 2.0 to version 2.1.

$\bullet$~Higgs production has been added, which implies new process-specific 
files\\ ({\code mcatnlo\_hgmain.f}, {\code mcatnlo\_hgxsec.f}, 
{\code hgscblks.h}, {\code mcatnlo\_hwanhgg.f}), and a modification to
{\code mcatnlo\_hwlhin.f}. 

$\bullet$~Post-1999 PDF sets have been added to the MC@NLO PDF library.

$\bullet$~Script variables have been added to {\code MCatNLO.inputs}. 
Most of them are only relevant to Higgs production, and don't affect processes
implemented in version 2.0. One of them ({\variable LAMBDAHERW}) may affect
all processes: in version 2.1, the variables {\variable LAMBDAFIVE} and
{\variable LAMBDAHERW} are used to set the value of $\Lambda_{\sss QCD}$ in
NLO and MC runs respectively, whereas in version 2.0 {\variable LAMBDAFIVE}
controlled both. The new setup is necessary since modern PDF sets have
$\Lambda_{\sss QCD}$ values which are too large to be supported by \HW.
(Recall that the effect of using {\variable LAMBDAHERW} different from
{\variable LAMBDAFIVE} is beyond NLO.)

$\bullet$~The new script variable {\variable PDFPATH} should be set equal 
to the name of the directory where the PDF grid files (which can be downloaded
from the MC@NLO web page) are stored. At run time, when executing {\code
runNLO}, or {\code runMC}, or {\code runMCatNLO}, logical links to these
files will be created in the running directory (in version 2.0, this
operation had to be performed by the user manually).

$\bullet$~Minor bugs corrected in {\code mcatnlo\_hbook.f} and sample 
analysis routines.

\subsection{From MC@NLO version 2.1 to version 2.2\label{app:newverb}}
In this appendix we list the changes that occurred in the package
from version 2.1 to version 2.2.

$\bullet$~Single vector boson production has been added, which implies 
new process-specific files ({\code mcatnlo\_sbmain.f}, 
{\code mcatnlo\_sbxsec.f}, {\code svbcblks.h}, {\code mcatnlo\_hwansvb.f}), 
and a modification to {\code mcatnlo\_hwlhin.f}. 

$\bullet$~The script variables {\variable WWIDTH} and {\variable ZWIDTH}
have been added to {\code MCatNLO.inputs}. These denote the physical widths 
of the $W$ and $Z^0$ bosons, used to generate the mass distributions of
the vector bosons according to the Breit--Wigner function, in the case
of single vector boson production (vector boson pair production is
still implemented only in the zero-width approximation).

\subsection{From MC@NLO version 2.2 to version 2.3\label{app:newverc}}
In this appendix we list the changes that occurred in the package
from version 2.2 to version 2.3.

$\bullet$~Lepton pair production has been added, which implies 
new process-specific files ({\code mcatnlo\_llmain.f}, 
{\code mcatnlo\_llxsec.f}, {\code llpcblks.h}, {\code mcatnlo\_hwanllp.f}), 
and modifications to {\code mcatnlo\_hwlhin.f} and 
{\code mcatnlo\_hwdriver.f}.

$\bullet$~The script variable {\variable AEMRUN} has been added, since
the computation of single vector boson and lepton pair cross sections is
performed in the $\MSbar$ scheme (the on-shell scheme was previously
used for single vector boson production).

$\bullet$~The script variables {\variable FRENMC} and {\variable FFACTMC}
have been eliminated. 

$\bullet$~The structure of pseudo-random number generation in heavy
flavour production has been changed, to avoid a correlation that
affected the azimuthal angle distribution for the products of the hard
partonic subprocesses.

$\bullet$~A few minor bugs have been corrected, which affected the rapidity
of the vector bosons in single vector boson production (a 2--3\% effect),
and the assignment of $\Lambda_{\sss QCD}$ for the LO and NLO PDF sets of
Alekhin.

\subsection{From MC@NLO version 2.3 to version 3.1\label{app:newverd}}
In this appendix we list the changes that occurred in the package
from version 2.3 to version 3.1.

$\bullet$~Associated Higgs production has been added, which implies 
new process-specific files ({\code mcatnlo\_vhmain.f}, 
{\code mcatnlo\_vhxsec.f}, {\code vhgcblks.h}, {\code mcatnlo\_hwanvhg.f}), 
and modifications to {\code mcatnlo\_hwlhin.f} and 
{\code mcatnlo\_hwdriver.f}.

$\bullet$~Spin correlations in $W^+W^-$ production and leptonic decay
have been added;
the relevant codes ({\code mcatnlo\_vpmain.f}, {\code mcatnlo\_vhxsec.f})
have been modified; the sample analysis routines ({\code mcatnlo\_hwanvbp.f})
have also been changed. Tree-level matrix elements have been computed with
MadGraph/MadEvent~\cite{Stelzer:1994ta,Maltoni:2002qb}, which uses 
HELAS~\cite{Murayama:1992gi}; the relevant routines and common blocks 
are included in {\code mcatnlo\_helas2.f} and {\code MEcoupl.inc}.

$\bullet$~The format of the event file has changed in several respects,
the most relevant of which is that the four-momenta are now given as
$(p_x,p_y,p_z,m)$ (up to version 2.3 we had $(p_x,p_y,p_z,E)$). Event
files generated with version 2.3 or lower {\em must not be used} with
version 3.1 or higher (the code will prevent the user from doing so).

$\bullet$~The script variables {\variable GAMMAX}, {\variable MASSINF},
and {\variable MASSSUP} have been replaced with {\variable xGAMMAX}, 
{\variable xMASSINF} and {\variable xMASSSUP}, with {\variable x=H,V1,V2}.

$\bullet$~New script variables {\variable IVCODE}, {\variable IL1CODE},
and {\variable IL2CODE} have been introduced.

$\bullet$~Minor changes have been made to the routines that put the partons
on the \HWs\ mass shell for lepton pair, heavy quark, and vector boson pair
production; effects are beyond the fourth digit.

$\bullet$~The default electroweak parameters have been changed for
vector boson pair production, in order to make them consistent with those
used in other processes. The cross sections are generally smaller
in version 3.1 wrt previous versions, the dominant effect being the 
value of $\sinthW$: we have now $\sinsqthW=0.2311$, in lower
versions $\sinsqthW=1-m_W^2/m_Z^2$. The cross sections
are inversely proportional to $\sinfthW$.

\subsection{From MC@NLO version 3.1 to version 3.2\label{app:newvere}}
In this appendix we list the changes that occurred in the package
from version 3.1 to version 3.2.

$\bullet$~Single-$t$ production has been added, which implies 
new process-specific files\\ ({\code mcatnlo\_stmain.f}, 
{\code mcatnlo\_stxsec.f}, {\code stpcblks.h}, {\code mcatnlo\_hwanstp.f}), 
and modifications to {\code mcatnlo\_hwlhin.f} and 
{\code mcatnlo\_hwdriver.f}.

$\bullet$~LHAPDF library is now supported, which implies modifications to
all {\code *main.f} files, and two new utility codes,
{\code mcatnlo\_lhauti.f} and {\code mcatnlo\_mlmtolha.f}.

$\bullet$~New script variables {\variable Vud}, {\variable LHAPATH},
and {\variable LHAOFL} have been introduced.

$\bullet$~A bug affecting Higgs production has been fixed, which implies
a modification to {\code mcatnlo\_hgxsec.f}. Cross sections change with
respect to version 3.1 {\em only if} {\variable FFACT}$\ne 1$ (by 
${\cal O}(1\%)$ in the range $1/2\le$ {\variable FFACT} $\le 2$).

\subsection{From MC@NLO version 3.2 to version 3.3\label{app:newverf}}
In this appendix we list the changes that occurred in the package
from version 3.2 to version 3.3.

$\bullet$~Spin correlations have been added to $t\bar{t}$ and single-$t$
production processes, which imply modifications to several codes
({\code mcatnlo\_qqmain.f}, {\code mcatnlo\_qqxsec.f},
{\code mcatnlo\_stmain.f}, {\code mcatnlo\_stxsec.f},
{\code mcatnlo\_hwlhin.f} and {\code mcatnlo\_hwdriver.f}).
Tree-level matrix elements have been computed with
MadGraph/MadEvent~\cite{Stelzer:1994ta,Maltoni:2002qb}.

$\bullet$~The matching between NLO matrix elements and parton shower
is now smoother in Higgs production, which helps eliminate one unphysical 
feature in the $\pt$ spectra of the accompanying jets. The code 
{\code mcatnlo\_hgmain.f} has been modified. Technical details on this
matching procedure will be posted on the MC@NLO web page.

$\bullet$~The new script variable {\variable TWIDTH} has been introduced.

$\bullet$~All instances of {\variable HWWARN('s',i,*n)} have been
replaced with {\variable HWWARN('s',i)} in \HWs-related codes. This
is consistent with the definition of {\variable HWWARN} in \HWs\ versions
6.510 and higher; the user must be careful if linking to \HWs\ versions,
in which the former form of {\variable HWWARN} is used. Although \HWs\ 6.510 
compiles with {\variable g95} or {\variable gfortran}, MC@NLO 3.3
does not.

\subsection{From MC@NLO version 3.3 to version 3.4\label{app:newverg}}
In this appendix we list the changes that occurred in the package
from version 3.3 to version 3.4.

$\bullet$~$Wt$ production has been implemented, which implies new 
process-specific codes ({\code mcatnlo\_wtmain\_dr.f},
{\code mcatnlo\_wtmain\_ds.f}, {\code mcatnlo\_wtxsec\_dr.f}\\ and
{\code mcatnlo\_wtxsec\_ds.f}).

$\bullet$~Owing to the implementation of $Wt$ production and of top
hadronic decays, the Les Houches interface ({\code mcatnlo\_hwlhin.f}) 
and the driver ({\code mcatnlo\_hwdriver.f}) have been upgraded.

$\bullet$~New script variables ({\variable BRTOPTOx} and 
{\variable BRWTOx}, with {\variable x=LEP,HAD}; {\variable yGAMMAX},
{\variable yMASSINF} and {\variable yMASSSUP} with {\variable y=T1,T2};
{\variable TOPDECAY}; {\variable WTTYPE}; {\variable PTVETO})
have been introduced.

$\bullet$~The new script variables {\variable EXTRALIBS}, 
{\variable EXTRAPATHS}, and {\variable INCLUDEPATHS} can be used to 
link to external libraries. Their use has only been tested on a recent 
Scientific Linux release, and they may be not portable to other systems.

$\bullet$~The ranges of variables {\variable ILxCODE} have been extended
for several processes, in order to account for the newly-implemented
hadronic decays.

$\bullet$~{\code MCatNLO.inputs} and {\code MCatNLO.Script} have been
upgraded to reflect the changes above. A new sample input file
({\code MCatNLO\_rb.inputs}) is included, which documents the
use of an analysis producing plots in Root format. Finally, the possibility
is given to link to a dynamic LHAPDF library (through 
{\code MCatNLO\_dyn.Script} and {\code Makefile\_dyn}).

$\bullet$~Front-end Fortran routines ({\code rbook\_fe.f}) are provided,
to produce plots in Root format, using the same syntax as for calling our
HBOOK-type routines. A companion C++ code is needed ({\code rbook\_be.cc}).
These codes have been written by W.~Verkerke. Examples of 
analysis routines using Root format have been added 
({\code mcatnlo\_hwan{\em xxx}\_rb.f}). A call to a release-memory
routine ({\code RCLOS}) has been added to {\code mcatnlo\_hwdriver.f}; 
this is only needed when using a Root-format output, and a dummy 
body of {\code RCLOS} has been added to HBOOK-format analysis
files {\code mcatnlo\_hwan{\em xxx}.f}.

$\bullet$~The linking to LHAPDF has been upgraded, assuming the use
of LHAPDF version 5.0 or higher. The file {\code mcatnlo\_lhauti.f}
has been eliminated, and replaced with {\code mcatnlo\_utilhav4.f},
which is however necessary only if the user wants to link with
LHAPDF versions 4.xx (in such a case, the user will also need to
edit the Makefile).

$\bullet$~The automatic assignment of $\Lambda_{\sss QCD}$ when
using LHAPDF is now to be considered robust. This implies changes
to {\code mcatnlo\_mlmtolha.f}, the insertion of a dummy routine
into {\code mcatnlo\_mlmtopdf.f} and {\code mcatnlo\_pdftomlm.f},
and very minor changes to all {\code *main*.f} files.

$\bullet$~Minor changes to {\code mcatnlo\_hbook.f}, mainly affecting
two-dimensional plot outputs.

$\bullet$~A bug has been fixed, which prevented one from choosing properly 
the $W$ mass ranges in $W^+W^-$ production and subsequent decays in
the case of {\variable ViGAMMAX}$<0$ (thanks to F.~Filthaut).

$\bullet$~A bug has been fixed, which affected the computation of 
branching ratios in $t\bar{t}$ and single-top production; $\alpha_{em}(q^2)$
was previously called with argument $m_{top}$ rather than $m_{top}^2$.
This only affects event weights (i.e. not distributions), and is
numerically very small.

$\bullet$~A bug in \HW\ versions 6.500 -- 6.510 can lead to occasional
violation of momentum conservation when the \HWs\ parameter
{\code PRESPL=.FALSE.} (hard subprocess rapidity preserved), as is
formally assumed in MC@NLO.  Therefore at present we leave this
parameter at its default value, {\code PRESPL=.TRUE.} (hard subprocess 
longitudinal momentum preserved).  We have checked that  this formal
inconsistency has negligible actual consequences.  The bug has been
fixed in \HWs\ version 6.520. With older versions, the fix may be found 
on the Fortran \HW\ wiki at
http://projects.hepforge.org/fherwig/trac/report (ticket 33).  When
this fix is implemented, the statement {\code PRESPL=.FALSE.} must be
inserted in {\code mcatnlo\_hwdriver.f} at the place indicated by the
comments therein.

$\bullet$~It has been found that a simpler form for the MC subtraction
terms with respect to that of eq.~(B.43) of ref.~\cite{Frixione:2003ei}
can be adopted; this form is now implemented in version 3.4.
This change is relevant only to $Q\bar{Q}$ and single-top
production, since for the other processes the new form and that of 
eq.~(B.43) (which is implemented in MC@NLO version 3.3 or earlier)
coincide. The differences between the two forms are equivalent 
to power-suppressed terms. This has been verified 
by comparing results obtained with version 3.4 for $t\bar{t}$ 
and single-top ($s$- and $t$-channel) production at the Tevatron and 
the LHC, and for $b\bar{b}$ production at the Tevatron, with analogous
results obtained with version 3.3. On the other hand, $b\bar{b}$ production 
at the LHC does display large differences, owing to the fact that 
the old form of MC subtraction terms has a pathology which affects
this process. Starting from version 3.4 $b\bar{b}$ production 
at the LHC may be considered safe. Technical details on the new form
of the MC subtraction terms will be posted on the MC@NLO web page.

\subsection{From MC@NLO version 3.4 to version 4.0\label{app:newverh}}
In this appendix we list the changes that occurred in the package
from version 3.4 to subversions 3.41 and 3.42 to version 4.0.

$\bullet$~A problem was found which affected top decays in the
processes listed in table~\ref{tab:procdec} (except for $H^\pm t$, not
implemented in version 3.4). This implied that the identities of 
top decay products in $n$ event samples of $k$ events each could have 
been statistically not equivalent to those of one single event sample 
of $n\times k$ events, for $k\simeq 5000$ or smaller. Fixed in
subversion 3.41.

$\bullet$~All processes have been interfaced to \HWpp (except for
$H^\pm t$ production). This implies a new structure of the source 
directory, and the addition of scripts ({\code MCatNLO\_pp.Script}) 
and a makefile ({\code Makefile\_pp}) specific to \HWpp.

$\bullet$~The linking to the static or dynamic LHAPDF library is now
done via a shell variable. {\code MCatNLO\_dyn.Script} and 
{\code Makefile\_dyn} are thus obsolete, and have been removed from 
the package.

$\bullet$~$H^\pm t$ production has been implemented, including
spin correlations.

$\bullet$~Spin correlations and anomalous couplings have been included
in $W^\pm Z$ production.

$\bullet$~A numerical inaccuracy problem which affected the large-rapidity, 
large-$\pt$ region of leptons in $W$ production has been fixed in
subversion 3.42. In addition, the convention on the range of lepton-parton 
azimuthal angular differences has been changed from $(0,\pi)$ to
$(-\pi,\pi)$ in this process.

$\bullet$~The values of the fractions of the longitudinal momenta of
the incoming partons $x_1$ and $x_2$, and that of the mass scale squared
$Q^2$ (in GeV$^2$), used in the computations of the PDFs, are now stored
in the event file.

\section{Running the package without the shell scripts\label{app:instr}}
In this appendix, we describe the actions that the user needs to 
take in order to run the package without using the shell scripts,
and the {\variable Makefile}. Examples are given for vector boson
pair production, but only trivial modifications are necessary in
order to treat other production processes.

\subsection{Creating the executables\label{app:exe}}
An $\MCatNLO$ run requires the creation of two executables, for the \NLO\
and \MC\ codes respectively. The files to link depend on whether one
uses PDFLIB, LHAPDF, or the PDF library provided with this package; 
we list them below:
\begin{itemize}
\item {\bf NLO with private PDFs:}
{\code mcatnlo\_vbmain.o mcatnlo\_vbxsec.o mcatnlo\_helas2.o 
mcatnlo\_date.o mcatnlo\_int.o mcatnlo\_uxdate.o mcatnlo\_uti.o 
mcatnlo\_str.o mcatnlo\_pdftomlm.o mcatnlo\_libofpdf.o dummies.o SYSFILE}
\item {\bf NLO with PDFLIB:}
{\code mcatnlo\_vbmain.o mcatnlo\_vbxsec.o mcatnlo\_helas2.o 
mcatnlo\_date.o mcatnlo\_int.o mcatnlo\_uxdate.o mcatnlo\_uti.o 
mcatnlo\_str.o mcatnlo\_mlmtopdf.o dummies.o}
{\variable SYSFILE CERNLIB}
\item {\bf NLO with LHAPDF:}
{\code mcatnlo\_vbmain.o mcatnlo\_vbxsec.o mcatnlo\_helas2.o 
mcatnlo\_date.o mcatnlo\_int.o mcatnlo\_uxdate.o mcatnlo\_lhauti.o 
mcatnlo\_str.o mcatnlo\_mlmtolha.o dummies.o}
{\variable SYSFILE LHAPDF}
\item {\bf MC with private PDFs:}
{\code mcatnlo\_hwdriver.o mcatnlo\_hwlhin.o mcatnlo\_hwanvbp.o 
mcatnlo\_hbook.o mcatnlo\_str.o mcatnlo\_pdftomlm.o mcatnlo\_libofpdf.o 
dummies.o} {\variable HWUTI HERWIGVER}
\item {\bf MC with PDFLIB:}
{\code mcatnlo\_hwdriver.o mcatnlo\_hwlhin.o mcatnlo\_hwanvbp.o 
mcatnlo\_hbook.o mcatnlo\_str.o mcatnlo\_mlmtopdf.o dummies.o}
{\variable HWUTI HERWIGVER CERNLIB}
\item {\bf MC with LHAPDF:}
{\code mcatnlo\_hwdriver.o mcatnlo\_hwlhin.o mcatnlo\_hwanvbp.o 
mcatnlo\_hbook.o mcatnlo\_str.o mcatnlo\_mlmtolha.o dummies.o}
{\variable HWUTI HERWIGVER LHAPDF}
\end{itemize}
The process-specific codes {\code mcatnlo\_vbmain.o} and
{\code mcatnlo\_vbxsec.o} (for the \NLO\ executable) and
{\code mcatnlo\_hwanvbp.o} (the \HWs\ analysis routines in the
\MC\ executable) need to be replaced by their analogues for 
other production processes.

The variable {\variable SYSFILE} must be set either equal to {\code alpha.o},
or to {\code linux.o}, or to {\code sun.o}, according to the architecture 
of the machine on which the run is performed. For any other architecture,
the user should provide a file corresponding to {\code alpha.f} etc.,
which he/she will easily obtain by modifying {\code alpha.f}. The 
variables {\variable HWUTI} and {\variable HERWIGVER} have been described
in sect.~\ref{sec:scrvar}. In order to create the object files eventually 
linked, static compilation is always recommended (for example, 
{\code g77 -Wall -fno-automatic} on Linux).

\subsection{The input files\label{app:input}}
Here, we describe the inputs to be given to the \NLO\ and 
\MC\ executables in the case of vector boson pair production. The case
of other production processes is completely analogous.
When the shell scripts are used to run the $\MCatNLO$,
two files are created, {\variable FPREFIXNLOinput} and 
{\variable FPREFIXMCinput}, which are read by the \NLO\ and \MC\ executable
respectively. We start by considering the inputs for the \NLO\
executable, presented in table~\ref{tab:NLOi}.
\begin{table}[htb]
\begin{center}
\begin{tabular}{ll}
\hline
 '{\variable FPREFIX}'                       & ! prefix for BASES files\\
 '{\variable EVPREFIX}'                      & ! prefix for event files\\
  {\variable ECM FFACT FREN FFACTMC FRENMC}  & ! energy, scalefactors\\
  {\variable IPROC}                        & ! -2850/60/70/80=WW/ZZ/ZW+/ZW-\\
  {\variable WMASS ZMASS}                    & ! M\_W, M\_Z\\
  {\variable UMASS DMASS SMASS CMASS BMASS GMASS} & ! quark and gluon masses\\
 '{\variable PART1}'  '{\variable PART2}'    & ! hadron types\\
 '{\variable PDFGROUP}'   {\variable PDFSET} & ! PDF group and id number\\
  {\variable LAMBDAFIVE}                     & ! Lambda\_5, $<$0 for default\\
 '{\variable SCHEMEOFPDF}'                   & ! scheme\\
  {\variable NEVENTS}                        & ! number of events\\
  {\variable WGTTYPE}                 & ! 0 =$>$ wgt=+1/-1, 1 =$>$ wgt=+w/-w\\
  {\variable RNDEVSEED}                      & ! seed for rnd numbers\\
  {\variable zi}                             & ! zi\\
  {\variable nitn$_1$ nitn$_2$}              & ! itmx1,itmx2\\
\hline\\
\end{tabular}
\end{center}
\caption{\label{tab:NLOi}
Sample input file for the \NLO\ code (for vector boson pair production). 
{\variable FPREFIX} and {\variable EVPREFIX} must be understood with 
{\variable SCRTCH} in front (see sect.~\ref{sec:scrvar}).
}
\end{table}
The variables whose name is in uppercase characters have been described 
in sect.~\ref{sec:scrvar}. The other variables are assigned by the shell
script. Their default values are given in table~\ref{tab:defNLO}.
\begin{table}[htb]
\begin{center}
\begin{tabular}{ll}
\hline
Variable & Default value\\
\hline
{\variable zi}          & 0.2\\
{\variable nitn$_i$}    & 10/0 ({\variable BASES=ON/OFF})\\
\hline\\
\end{tabular}
\end{center}
\caption{\label{tab:defNLO}
Default values for script-generated variables in {\code FPREFIXNLOinput}.
}
\end{table}
Users who run the package without the script should use the values
given in table~\ref{tab:defNLO}. The variable {\variable zi} controls,
to a certain extent, the number of negative-weight events generated 
by the $\MCatNLO$ (see ref.~\cite{Frixione:2002ik}). Therefore, the user
may want to tune this parameter in order to reduce as much as possible
the number of negative-weight events. We stress that the \MC\ code will
not change this number; thus, the tuning can (and must) be done only 
by running the \NLO\ code. The variables {\variable nitn$_i$} control
the integration step (see sect.~\ref{sec:evfile}), which can be
skipped by setting {\variable nitn$_i=0$}. If one needs to perform the
integration step, we suggest setting these variables as indicated in
table~\ref{tab:defNLO}. 

\begin{table}[htb]
\begin{center}
\begin{tabular}{ll}
\hline
 '{\variable EVPREFIX.events}'               & ! event file\\
  {\variable NEVENTS}                        & ! number of events\\
  {\variable pdftype}                      & ! 0-$>$Herwig PDFs, 1 otherwise\\
 '{\variable PART1}'  '{\variable PART2}'    & ! hadron types\\
  {\variable beammom beammom}                & ! beam momenta\\
  {\variable IPROC}                         & ! --2850/60/70/80=WW/ZZ/ZW+/ZW-\\
 '{\variable PDFGROUP}'                      & ! PDF group (1)\\
  {\variable PDFSET}                         & ! PDF id number (1)\\
 '{\variable PDFGROUP}'                      & ! PDF group (2)\\
  {\variable PDFSET}                         & ! PDF id number (2)\\
  {\variable LAMBDAHERW}                     & ! Lambda\_5, $<$0 for default\\
  {\variable WMASS WMASS ZMASS}              & ! M\_W+, M\_W-, M\_Z\\
  {\variable UMASS DMASS SMASS CMASS BMASS GMASS} & ! quark and gluon masses\\
\hline\\
\end{tabular}
\end{center}
\caption{\label{tab:MCi}
Sample input file for the \MC\ code (for vector boson pair production), 
resulting from setting {\variable HERPDF=EXTPDF}, which implies 
{\variable pdftype=1}. 
Setting {\variable HERPDF=DEFAULT} results in an analogous file, with
{\variable pdftype=0}, and without the lines concerning
{\variable PDFGROUP} and {\variable PDFSET}. {\variable EVPREFIX} 
must be understood with {\variable SCRTCH} in front 
(see sect.~\ref{sec:scrvar}). The negative sign of {\variable IPROC}
tells the \EvG\ to use Les Houches interface routines.
}
\end{table}
We now turn to the inputs for the \MC\ executable, presented
in table~\ref{tab:MCi}. 
The variables whose names are in uppercase characters have been described 
in sect.~\ref{sec:scrvar}. The other variables are assigned by the shell
script. Their default values are given in table~\ref{tab:defMC}.
\begin{table}[htb]
\begin{center}
\begin{tabular}{ll}
\hline
Variable & Default value\\
\hline
{\variable esctype}         & 0\\
{\variable pdftype}         & 0/1 ({\variable HERPDF=DEFAULT/EXTPDF})\\
{\variable beammom}         & {\variable EMC}/2\\
\hline\\
\end{tabular}
\end{center}
\caption{\label{tab:defMC}
Default values for script-generated variables in {\code MCinput}.
}
\end{table}
The user can freely change the values of {\variable esctype} and
{\variable pdftype}; on the other hand, the value of {\variable beammom}
must always be equal to half of the hadronic CM energy.

When LHAPDF is linked, the value of {\variable PDFSET} is sufficient
to identify the parton density set. In such a case, {\variable PDFGROUP}
must be set in input equal to {\variable LHAPDF} if the user wants
to freeze the PDFs at the boundaries (defined as the ranges in which
the fits have been performed). If one chooses to extrapolate the PDFs
across the boundaries, one should set {\variable PDFGROUP=LHAEXT}
in input.

In the case of $\gamma/Z$, $W^\pm$, Higgs or heavy quark production, the 
\MC\ executable can be run with the corresponding positive input process 
codes {\variable IPROC} = 1350, 1399, 1499, 1600+ID, 1705, 1706,
2000--2008, 2600+ID or 2700+ID,
to generate a standard \HWs\ run for comparison purposes\footnote{For
vector boson pair production, for historical reasons, the different
process codes 2800--2825 must be used.}.  Then the input
event file will not be read: instead, parton configurations will be
generated by \HWs\ according to the LO matrix elements.



\begin{thebibliography}{100}

\bibitem{Frixione:2002ik}
S.~Frixione and B.~R.~Webber,
``Matching NLO QCD computations and parton shower simulations,''
JHEP {\bf 0206} (2002) 029
[hep-ph/0204244].

\bibitem{Frixione:2003ei}
S.~Frixione, P.~Nason and B.~R.~Webber,
``Matching NLO QCD and parton showers in heavy flavour production,''
JHEP {\bf 0308} (2003) 007
[arXiv:hep-ph/0305252].

\bibitem{Frixione:1995ms}
S.~Frixione, Z.~Kunszt and A.~Signer,
``Three-jet cross sections to next-to-leading order,''
Nucl.\ Phys.\ B {\bf 467} (1996) 399
[arXiv:hep-ph/9512328].

\bibitem{Frixione:1997np}
S.~Frixione,
``A general approach to jet cross sections in QCD,''
Nucl.\ Phys.\ B {\bf 507} (1997) 295
[arXiv:hep-ph/9706545].

\bibitem{Mele:1990bq}
B.~Mele, P.~Nason and G.~Ridolfi,
``QCD Radiative Corrections To Z Boson Pair Production In Hadronic Collisions,''
Nucl.\ Phys.\ B {\bf 357} (1991) 409.

\bibitem{Frixione:1992pj}
S.~Frixione, P.~Nason and G.~Ridolfi,
``Strong corrections to W Z production at hadron colliders,''
Nucl.\ Phys.\ B {\bf 383} (1992) 3.

\bibitem{Frixione:1993yp}
S.~Frixione,
``A Next-to-leading order calculation of the cross-section for the production of W+ W- pairs in hadronic collisions,''
Nucl.\ Phys.\ B {\bf 410} (1993) 280.

\bibitem{Dixon:1999di}
  L.~J.~Dixon, Z.~Kunszt and A.~Signer,
  ``Vector boson pair production in hadronic collisions at order $\alpha_s$ :
  Lepton correlations and anomalous couplings'',
  Phys.\ Rev.\  D {\bf 60} (1999) 114037
  [arXiv:hep-ph/9907305].

\bibitem{Mangano:1991jk}
  M.~L.~Mangano, P.~Nason and G.~Ridolfi,
  ``Heavy quark correlations in hadron collisions at next-to-leading order,''
  Nucl.\ Phys.\ B {\bf 373} (1992) 295.

\bibitem{Dawson:1990zj}
S.~Dawson,
``Radiative Corrections To Higgs Boson Production,''
Nucl.\ Phys.\ B {\bf 359} (1991) 283.

\bibitem{Djouadi:1991tk}
A.~Djouadi, M.~Spira and P.~M.~Zerwas,
``Production of Higgs bosons in proton colliders: QCD corrections,''
Phys.\ Lett.\ B {\bf 264} (1991) 440.

\bibitem{Altarelli:1979ub}
G.~Altarelli, R.~K.~Ellis and G.~Martinelli,
``Large Perturbative Corrections To The Drell-Yan Process In QCD,''
Nucl.\ Phys.\ B {\bf 157} (1979) 461.

\bibitem{Aurenche:1980tp}
P.~Aurenche and J.~Lindfors,
``QCD Corrections To Direct Lepton Production In Hadronic Collisions,''
Nucl.\ Phys.\ B {\bf 185} (1981) 274.

\bibitem{Oleari:2005inprep}
 V.~Del~Duca, S.~Frixione and B.~R.~Webber,
``MC@NLO for Higgs Boson Production,'' in preparation.

\bibitem{Harris:2002md}
  B.~W.~Harris, E.~Laenen, L.~Phaf, Z.~Sullivan and S.~Weinzierl,
  ``The fully differential single top quark cross section in  next-to-leading
  order QCD,''
  Phys.\ Rev.\ D {\bf 66} (2002) 054024
  [arXiv:hep-ph/0207055].

\bibitem{Weydert:2009vr}
  C.~Weydert {\it et al.},
  Eur.\ Phys.\ J.\  C {\bf 67} (2010) 617
  [arXiv:0912.3430 [Unknown]].

\bibitem{Giele:1995kr}
  W.~T.~Giele, S.~Keller and E.~Laenen,
  ``QCD Corrections to W Boson plus Heavy Quark Production at the Tevatron,''
  Phys.\ Lett.\  B {\bf 372} (1996) 141
  [arXiv:hep-ph/9511449].

\bibitem{Marchesini:1992ch}
G.~Marchesini, B.~R.~Webber, G.~Abbiendi, I.~G.~Knowles, M.~H.~Seymour and L.~Stanco,
``HERWIG: A Monte Carlo event generator for simulating hadron emission reactions with interfering gluons. Version 5.1 - April 1991,''
Comput.\ Phys.\ Commun.\  {\bf 67} (1992) 465.

\bibitem{Corcella:2001bw}
G.~Corcella, I.G.~Knowles, G.~Marchesini, S.~Moretti, K.~Odagiri,
P.~Richardson, M.H.~Seymour and B.R.~Webber,
``HERWIG 6: An event generator for hadron emission reactions with  interfering gluons (including supersymmetric processes),''
JHEP {\bf 0101} (2001) 010
[hep-ph/0011363].

\bibitem{Corcella:2002jc}
G.~Corcella {\it et al.},
``HERWIG 6.5 release note,''
arXiv:hep-ph/0210213.

\bibitem{Bahr:2008pv}
  M.~Bahr {\it et al.},
  ``Herwig++ Physics and Manual,''
  Eur.\ Phys.\ J.\  C {\bf 58} (2008) 639
  [arXiv:0803.0883 [hep-ph]].

\bibitem{Bahr:2008tf}
  M.~Bahr {\it et al.},
  ``Herwig++ 2.3 Release Note,''
  arXiv:0812.0529 [hep-ph].

\bibitem{Torrielli:2010aw}
  P.~Torrielli and S.~Frixione,
  ``Matching NLO QCD computations with PYTHIA using MC@NLO'',
  JHEP {\bf 1004} (2010) 110
  [arXiv:1002.4293 [Unknown]].

\bibitem{HWpppaper}
 S.~Frixione, F.~Stoeckli, P.~Torrielli, B.~R.~Webber, 
 ``NLO QCD corrections in Herwig++ with MC@NLO'',
 arXiv:1010.0568 [hep-ph].

\bibitem{Frixione:2005vw}
  S.~Frixione, E.~Laenen, P.~Motylinski and B.~R.~Webber,
  ``Single-top production in MC@NLO,''
  JHEP {\bf 0603} (2006) 092
  [arXiv:hep-ph/0512250].

\bibitem{Frixione:2008yi}
  S.~Frixione, E.~Laenen, P.~Motylinski, B.~R.~Webber and C.~D.~White,
  ``Single-top hadroproduction in association with a W boson,''
  JHEP {\bf 0807} (2008) 029
  [arXiv:0805.3067 [hep-ph]].

\bibitem{SFAOH}
 S.~Frixione, A~Oh,
 ``Anomalous couplings in $WZ$ hadroproduction'',
 in preparation.

\bibitem{Boos:2001cv}
E.~Boos {\it et al.},
``Generic user process interface for event generators,''
arXiv:hep-ph/0109068.

\bibitem{Frixione:2007zp}
  S.~Frixione, E.~Laenen, P.~Motylinski and B.~R.~Webber,
  ``Angular correlations of lepton pairs from vector boson and top quark decays
  JHEP {\bf 0704} (2007) 081
  [arXiv:hep-ph/0702198].

\bibitem{WVroot}
  W.~Verkerke, Nikhef, unpublished.

\bibitem{Lonnblad:2006pt}
  L.~Lonnblad,
  ``ThePEG, Pythia7, herwig++ and Ariadne'',
  Nucl.\ Instrum.\ Meth.\  A {\bf 559} (2006) 246.
  Web page: {\tt http://www.thep.lu.se/ThePEG}.

\bibitem{Whalley:2005nh}
  M.~R.~Whalley, D.~Bourilkov and R.~C.~Group,
  ``The Les Houches accord PDFs (LHAPDF) and LHAGLUE,''
  arXiv:hep-ph/0508110.

\bibitem{Kawabata:1995th}
S.~Kawabata,
``A New version of the multidimensional integration and event generation package BASES/SPRING,''
Comput.\ Phys.\ Commun.\  {\bf 88} (1995) 309.

\bibitem{White:2009yt}
  C.~D.~White, S.~Frixione, E.~Laenen and F.~Maltoni,
  ``Isolating Wt production at the LHC'',
  JHEP {\bf 0911} (2009) 074
  [arXiv:0908.0631 [hep-ph]].

\bibitem{Stelzer:1994ta}
  T.~Stelzer and W.~F.~Long,
  ``Automatic generation of tree level helicity amplitudes,''
  Comput.\ Phys.\ Commun.\  {\bf 81} (1994) 357
  [arXiv:hep-ph/9401258].

\bibitem{Maltoni:2002qb}
  F.~Maltoni and T.~Stelzer,
  ``MadEvent: Automatic event generation with MadGraph,''
  JHEP {\bf 0302} (2003) 027
  [arXiv:hep-ph/0208156].

\bibitem{Murayama:1992gi}
  H.~Murayama, I.~Watanabe and K.~Hagiwara,
  ``HELAS: HELicity amplitude subroutines for Feynman diagram evaluations,''
KEK-91-11

\end{thebibliography}
\end{document}